\documentclass[10pt,nofootinbib,twocolumn,superscriptaddress,aps,prd,longbibliography]{revtex4-2}
\usepackage{amsmath}
\usepackage{amssymb}
\usepackage{amsfonts}
\usepackage{soul}
\usepackage{hyperref}
\usepackage{color}
\usepackage{graphicx}

\usepackage{tikz,xcolor}
\definecolor{lime}{HTML}{A6CE39}
\DeclareRobustCommand{\orcidicon}{%
	\begin{tikzpicture}
	\draw[lime, fill=lime] (0,0) 
	circle [radius=0.16] 
	node[white] {{\fontfamily{qag}\selectfont \tiny ID}};
	\draw[white, fill=white] (-0.0625,0.095) 
	circle [radius=0.007];
	\end{tikzpicture}
	\hspace{-2mm}
}
\foreach \x in {A, ..., Z}{%
	\expandafter\xdef\csname orcid\x\endcsname{\noexpand\href{https://orcid.org/\csname orcidauthor\x\endcsname}{\noexpand\orcidicon}}
}
\begin{document}
%%%%%%%%%%%%%%%%%%%%%%%%%%%%%%%%
%%%%%%%%%%%%%%%%%%%%%%%%%%%%%%%%
%%%%%%%%%%%%%%%%%%%%%%%%%%%%%%%%
\date{\today}
\title{Spacetime effects on wavepackets of coherent light}

\author{David Edward Bruschi\orcidA{}}
\affiliation{Institute for Quantum Computing Analytics (PGI-12), Forschungszentrum J\"ulich, 52425 J\"ulich, Germany}
%\affiliation{Theoretical Physics, Universität des Saarlandes, 66123 Saarbrücken, Germany}
\email{david.edward.bruschi@posteo.net}
\author{Symeon Chatzinotas\orcidB{}}
\affiliation{Interdisciplinary Centre for Security, Reliability and Trust (SnT), University of Luxembourg, L-1855 Luxembourg}
\author{Frank K. Wilhelm\orcidC{}}
\affiliation{Institute for Quantum Computing Analytics (PGI-12), Forschungszentrum J\"ulich, 52425 J\"ulich, Germany}
%\affiliation{Theoretical Physics, Universität des Saarlandes, 66123 Saarbrücken, Germany}
\email{david.edward.bruschi@posteo.net}
\author{Andreas Wolfgang Schell\orcidD{}}
\affiliation{Institut f\"ur Festk\"orperphysik, Leibniz Universit\"at Hannover, 30167 Hannover, Germany}
\affiliation{Physikalisch-Technische Bundesanstalt, 38116 Braunschweig, Germany}

\begin{abstract}
We investigate the interplay between gravity and the quantum coherence present in the state of a pulse of light propagating in curved spacetime. We first introduce an operational way to distinguish between the overall shift in the pulse wavepacket and its genuine deformation after propagation. 
We then apply our technique to quantum states of photons that are coherent in the frequency degree of freedom, as well as to states of completely incoherent light. We focus on Gaussian profiles and frequency combs and find that the quantum coherence initially present can enhance the deformation induced by propagation in a curved background. These results further support the claim that genuine quantum features, such as quantum coherence, can be used to probe the  gravitational properties of physical systems. We specialize our techniques to Earth-to-satellite communication setups, where the effects of gravity are weak but can be tested with current satellite technologies.
\end{abstract}

\maketitle

%---------------------------------------------------------------------------------------------------%
\section*{Introduction}
%---------------------------------------------------------------------------------------------------%
Photons and pulses of light are physical systems that are central to many areas of science. The ability to engineer, transmit, detect, and manipulate light is key to successful experimental, theoretical and technological endeavours \cite{MacFarlane:Dowling:2003,Flamini:Spagnolo:2018,Slussarenko:Pryde:2019}. Modelling light can be done using classical electrodynamics alone, as well as quantum mechanics or quantum electrodynamics. The latter avenue has given rise to the field of quantum optics, which is now widely studied theoretically and employed experimentally \cite{Scully:Zubairy:1997}. 
Development of quantum optics has led to the formation of the rapidly growing area of quantum optical technologies, with applications such as 
quantum sensing~\cite{Rosi:Sorrentino:2014,Tartaglia:2016,Degen2017}, quantum computing~\cite{Obrien2007}, and quantum communication with conventional~\cite{Gisin2007,Vallone:Bacco:2015,Dequal:Vallone:2016,Vallone:Dequal:2016,Liao:Cai:2017,Yin:Cao:2017:v2,Liao:Yong:2017,Yin:Cao:2017,Namazi:Vallone:2017,Liao:Cai:2018,Han:Yong:2018,Calderaro:Agnesi:2018,Xu:Ma:2019,Dai:Shen:2020,Dequal:Trigo:2021} and small satellites~\cite{Oi:Ling:2017,Kodheli:Lagunas:2020}. These applications have in common the fact that, in order to surpass their classical counterparts, a tremendous control over quantum states is required. For quantum communication in particular, the quantum states to be employed need to be transmitted over large distances. A natural starting point is to use optical fibres, which unfortunately limit achievable distances to few-hundreds of kilometers \cite{Liu:Jiang:2021,Chen:Zhang:2021}. The alternative to this is to move part of the infrastructure to space, where satellites in orbit are used to generate, receive or relay signals \cite{Vallone:Bacco:2015,Dequal:Vallone:2016,Vallone:Dequal:2016,Liao:Cai:2017,Yin:Cao:2017:v2,Liao:Yong:2017,Yin:Cao:2017,Namazi:Vallone:2017,Oi:Ling:2017,Liao:Cai:2018,Han:Yong:2018,Calderaro:Agnesi:2018,Khan2018,Agnesi:Calderaro:2019,Xu:Ma:2019,Kodheli:Lagunas:2020,Dai:Shen:2020,Dequal:Trigo:2021}. This latter approach brings into play novel advantages, such as significantly larger achievable distances \cite{Dequal:Vallone:2016}, as well as novel difficulties. Regardless of the advances within this avenue of research, the role of gravity on quantum information protocols has been ignored mainly because of the belief that its effects are negligible. Nevertheless, since space-based science occurs in an environment that is inherently within the remit of general relativity, it is important to answer the question of how to include gravity in these endeavours. 

\begin{figure}[ht!]
    \centering
    \includegraphics[width=\linewidth]{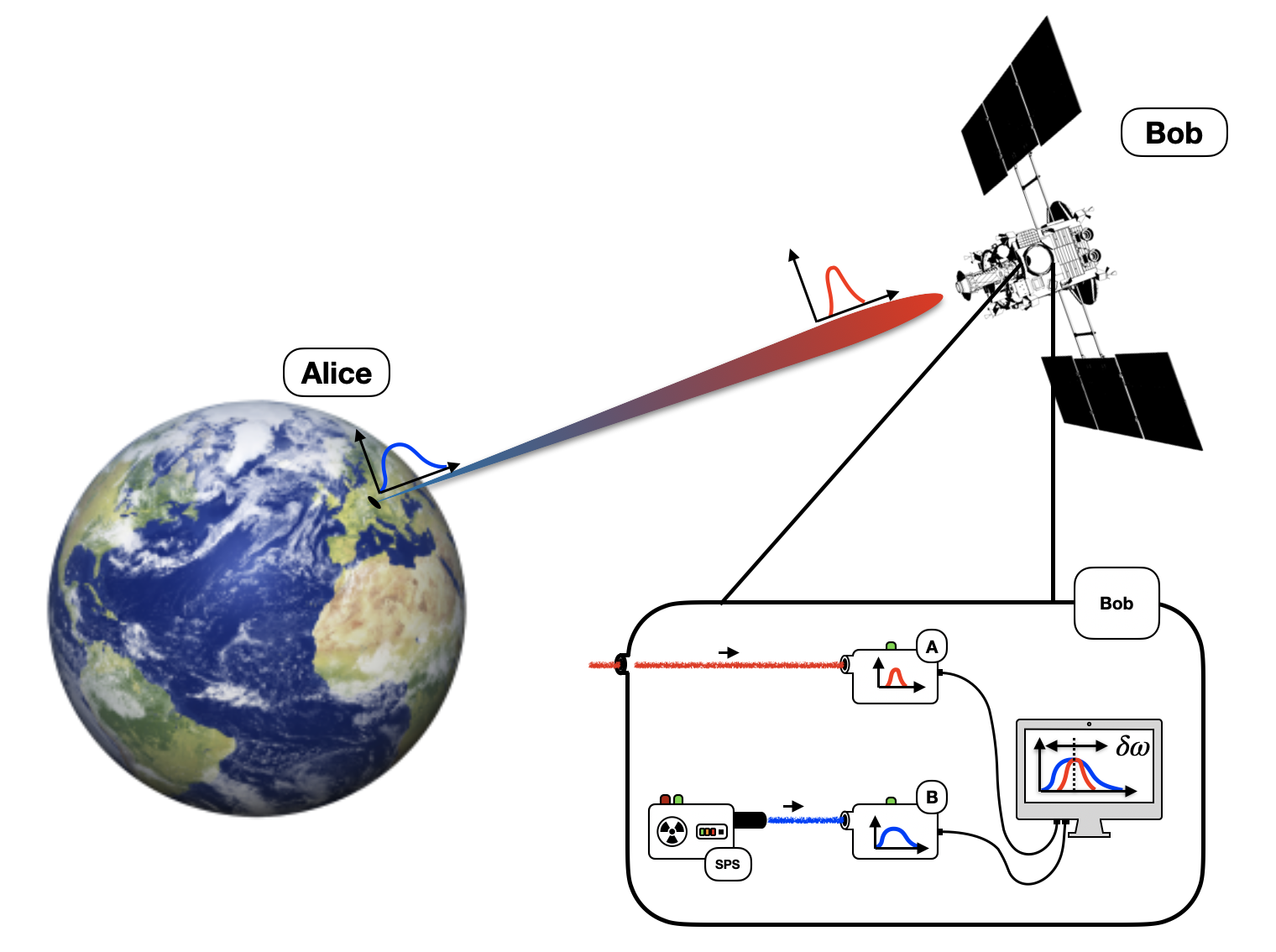}
    \caption{Alice sends photons to Bob. Photons are initially prepared with a chosen frequency distribution $F_{\omega_0}$ and are then detected by Bob as characterized by a different one $F'_{\omega_0'}$. This occurs due to gravitational redshift, and the relevant parameters, such as distance and mass of the planet, are now encoded in the received distribution $F'_{\omega_0'}$.}
    \label{fig0}
\end{figure}

The question of the effects of gravity on wavepackets of the electromagnetic field has been recently addressed in the context of quantum communications with satellite-based platforms \cite{Bruschi:Ralph:2014,Bruschi:Datta:2014,Kohlrus:Bruschi:2017}. Wavepackets of light were used to model a photon or a pulse of light travelling between two users located at different heights in the gravitational potential of the Earth. It was found that the receiver would detect a mismatch in the frequency distribution between the received wavepacket and the expected one, which encodes important information about relevant parameters, such as the distance between users, the mass of the Earth and its rotation \cite{Bruschi:Ralph:2014,Bruschi:Datta:2014,Kohlrus:Bruschi:2017}. Extension to include the effects of gravity on the polarization of the photons have also been performed \cite{Skrotskii:1957,Kohlrus:Louko:2019}. 

More recently, complementary work has quantified the distortion of the wavepacket in the degrees of freedom perpendicular to the path of propagation \cite{Exirifard:Culf:2021}. Together, this body of research provides a better understanding of the effects incurred by photons that propagate in curved spacetimes. 
The rapid advances in the study of satellite based quantum communication \cite{Vallone:Bacco:2015,Dequal:Vallone:2016,Vallone:Dequal:2016,Liao:Cai:2017,Yin:Cao:2017:v2,Liao:Yong:2017,Yin:Cao:2017,Namazi:Vallone:2017,Oi:Ling:2017,Liao:Cai:2018,Han:Yong:2018,Calderaro:Agnesi:2018,Khan2018,Agnesi:Calderaro:2019,Xu:Ma:2019,Kodheli:Lagunas:2020,Dai:Shen:2020,Dequal:Trigo:2021} make it necessary to obtain a comprehensive understanding of the effects of spacetime curvature on photons \cite{Kocsis:Loeb:2007}, given their core role as information carriers in future technologies. 

In this work we use quantum field theory in (weakly) curved spacetime to answer a fundamental question concerning the effects of gravity on propagating light: \textit{is it possible to separate and quantify the genuine distortion experienced by a wavepacket of light due to propagation on a curved background from the  the rigid shift of the wavepacket as a whole?}
To answer this question we define an operational way to quantify such genuine distortion, and we compute it for simple, yet experimentally realizable, wavepackets. We find that, while bell-shaped (i.e., Gaussian) packets without any complex phase are resilient to the deformation and witness small changes, as expected for such functions \cite{Kish:Ralph:2016}, wavepackets with complex phases experience distortions that can be potentially large, and can be detected with current technology.
Our results also highlight the difference inherent between initial states of light that are pure compared to initial states of light that are completely mixed. In particular, the coherence in the states, which is encoded in a relative phase of the sharp-frequency photonic states, can significantly enhance the genuine distortion. This reinforces the intuition that quantum coherence can influence gravitational properties of quantum systems.

We believe that the theoretical results of our work provide additional insight for the development of novel tests of quantum field theory in curved spacetime, as well as of advanced theories of Nature \cite{Kozameh:Parisi:2007,Rideout:Jennewein:2012,Tartaglia:2016,Xu:Ma:2019}. Finally, these results will also provide new ways to exploit the information encoded in the deformation of wavepackets in the development of novel technologies for sensing, positioning \cite{Calderaro:Agnesi:2018}, time-sharing \cite{Dai:Shen:2020} and communication \cite{Vallone:Bacco:2015,Liao:Cai:2017,Oi:Ling:2017,Kodheli:Lagunas:2020}.

This work is organized as follows. In Section \ref{section:tools} we introduce the necessary tools. In Section \ref{section:main} we provide our main operational tool to distinguish between deformation and translation of wavepackets of light. In Section \ref{section:applications} we apply our main proposal to concrete wavepackets. In Section \ref{section:discussion} we comment on the result of this work.

%\hl{Potential other titles}
%Gravitational effects on wavepackets of light OR\\
%Distinguishing between gravitational redshift and wavepacket distortions of light propagating in curved spacetime OR\\
%Spacetime curvature distorts wavepackets of light OR\\

%---------------------------------------------------------------------------------------------------%
\section{Propagation of photons in curved spacetime}\label{section:tools}
%---------------------------------------------------------------------------------------------------%
The propagation of light (in curved spacetime) is an important topic of research. An important aspect that is crucial to our work is that of the nature and interpretation of the effects of spacetime curvature on the defining quantities of a photon: its frequency and momentum. There have been many studies in this direction, which have come to ``different conclusions'' \cite{Okun:2000,Wilhelm:Bhola:2014}. While the formalism to quantify the redshift-like effects is not a topic of debate, the interpretation of the results is at least twofold: on the one hand, it has been suggested that energy and frequency of the photon are conserved, and that the redshift-like effects are due to changes in the energy levels of local atoms at the sender and receiver's location \cite{Okun:2000}. On the other, locally defined quantities along the null geodesic followed by a photon are not considered as the correct ones to be used, and a position-dependent frequency shift is obtained by defining frequency using a local clock at the each user's end. Regardless of the point of view taken, we stress that the overall physical conclusion is the same: \textit{experimentally measured} photons suffer from gravitational redshift \cite{Pound:Rebka:1959}.   

The effects of the (weakly) curved spacetime surrounding the Earth on the frequency profile of photons propagating between two users placed at different heights in the gravitational potential have been considered recently in order to provide realistic modelling of quantum communication setups \cite{Bruschi:Ralph:2014,Bruschi:Datta:2014,Kohlrus:Bruschi:2017}. More detailed analysis of propagation in arbitrary curved backgrounds has also been recently considered by computing including the distortion effects of curvature on propagating pulses of light in the dimensions \textit{orthogonal} to the path of propagation \cite{Exirifard:Culf:2021}. Here we will employ the original approach \cite{Bruschi:Ralph:2014,Bruschi:Datta:2014,Kohlrus:Bruschi:2017} and will assume that transversal deformations can be safely ignored. We leave it to further work to establish a complete quantitative estimation of the effects for concrete $3$-dimensional setups.

%---------------------------------------------------------------------------------------------------%
\subsection{Spacetime surrounding the Earth}
%---------------------------------------------------------------------------------------------------%
Light propagating in curved spacetime can be modelled, for simplicity, as scalar field $\hat{\varphi}(x^\mu)$ propagating through a classical curved background with metric $\boldsymbol{g}$ and spherical coordinates $(t,r,\theta,\phi)$. Such field is suitable to describe qualitatively one polarization of the electromagnetic field \cite{Friis:Lee:2013}. Distances in curved spacetime are obtained using the infinitesimal line element $ds^2:=g_{\mu\nu}dx^\mu dx^\nu$, which in turn is determined by the components $g_{\mu\nu}$ of the metric. In the case of the spacetime surrounding a (static) planet, such as the Earth, we can employ the Schwarzschild metric
\begin{align}\label{Schwarzschild:metric}
\mathbf{g}=\textrm{diag}(-f(r),f^{-1}(r),r^2,r^2\,\sin^2\theta)
\end{align}
with good approximation. Here we have defined $f(r):=1-(2\,M_\oplus)/r$ and $M_\oplus$ is the mass of the planet.\footnote{In this work we use units $\hbar=c=G=1$ unless explicitly states. This means that wherever the mass of the Earth $M_\oplus$ appears it should be replaced by $G_{\textrm{N}}\,M_\oplus/c^2$. In these unites, mass has dimension length. We employ Einstein's summation convention. } The metric \eqref{Schwarzschild:metric} can be complemented with boundary conditions at the Earth's surface $r=r_{\textrm{E}}$, but we note that we will focus on effects that do not depend on light impinging on the planet's surface and therefore ignore this issue. Rotation can also be included by extending the metric \eqref{Schwarzschild:metric} to the Kerr metric \cite{Misner:Thorne:1973}, but the effects of rotation are negligible \cite{Kohlrus:Bruschi:2017} to lowest order and we leave to its inclusion to future work.

The classical field $\varphi(x^\mu)$ satisfies the Klein-Gordon equation $\square\varphi:=(-g)^{-1/2}\partial_\mu((-g)^{1/2}g^{\mu\nu}\partial_\nu)\varphi=0$, where $g$ is the determinant of the metric. The quantum field can then be decomposed as $\hat{\varphi}(x^\mu)=\int d^3k [\hat{a}_{\boldsymbol{k}}\,u_{\boldsymbol{k}}(x^\mu)+\hat{a}^\dag_{\boldsymbol{k}}\,u^*_{\boldsymbol{k}}(x^\mu)]$ when there is a natural timelike Killing vector $\partial_\xi$ (i.e., a natural direction of time), and the modes $u_{\boldsymbol{k}}(x^\mu)$ satisfy $\square u_{\boldsymbol{k}}(x^\mu)=0$, as well as the eigenvalue equation $\partial_\xi u_{\boldsymbol{k}}=-i \omega_{\boldsymbol{k}} u_{\boldsymbol{k}}$ for appropriate field frequencies $\omega_{\boldsymbol{k}}$. The annihilation and creation operators $\hat{a}_{\boldsymbol{k}},\hat{a}^\dag_{\boldsymbol{k}}$ satisfy the canonical commutation relations $[\hat{a}_{\boldsymbol{k}},\hat{a}^\dag_{\boldsymbol{k}'}]=\delta^3(\boldsymbol{k}-\boldsymbol{k}')$, while all other commutators vanish.

In weakly curved spacetime we can work within the framework of linearized gravity, where $g_{\mu\nu}=\eta_{\mu\nu}+\epsilon h_{\mu\nu}$, $\boldsymbol{\eta}=\textrm{diag}(-1,1,1,1)$ is the Minkowski metric of flat spacetime, and $h_{\mu\nu}$ is the perturbation to Minkowski metric \cite{Misner:Thorne:1973}. Here, $\epsilon\ll1$ is a control parameter that we can assume upper bounded by $\epsilon\approx r_{\textrm{S}}/r_{\textrm{E}}\sim 1.4\times10^{-9}$, where $r_{\textrm{S}}:=2\,G_{\textrm{N}}M_\oplus/c^2\sim 9$~mm is the Schwarzschild radius associated to the Earth\footnote{A compact object that is contained completely within its Schwarzschild radius will collapse to form a black hole.}.

%---------------------------------------------------------------------------------------------------%
\subsection{Propagation of light}
%---------------------------------------------------------------------------------------------------%
We assume that we can model a photon as a wavepacket of the solutions $u_{\boldsymbol{k}}(x^\mu)$ to the field equations $\square u_{\boldsymbol{k}}(x^\mu)=0$, and that we can exclude all effects, such as dispersion and diffraction, suffered by the photon during propagation \cite{Bruschi:Ralph:2014}. We then assume that the photon is fundamentally confined in the direction of propagation, and therefore we neglect to first approximation the effects of the dimensions orthogonal to the propagation (extra dimensions can be included if necessary \cite{Exirifard:Culf:2021}). Other effects of propagation in curved spacetimes can also be studied and included \cite{Kocsis:Loeb:2007,Pedrosa:Rosas:2012,Raetzel:Wilkens:2016}. All together, these assumptions allow us to approximate the annihilation operator of a localized (in the propagation dimension) photon as
\begin{align}\label{photon:operator}
\hat{a}_{\omega_0}(t)=\int_0^{+\infty} d\omega\,e^{-i\,\omega\,(h(r)-t)}F_{\omega_0}(\omega)\,\hat{a}_\omega,
\end{align}
where $h(r)=h(r-r_0)$ is related to the coordinate distance $r-r_0$ travelled by the photon from position $r_0$ in flat spacetime \cite{Misner:Thorne:1973}. The function $F_{\omega_0}(\omega)$ defines the frequency profile of the photon, is peaked around $\omega_0$ with width $\sigma$ and is normalized by $\int d\omega |F_{\omega_0}(\omega)|^2=1$.

The photon operator \eqref{photon:operator} is constructed in a way such that, at time $t=0$, the photon is localized around height $r=r_0$, while it is localized at the height $r$ at time $t=h(r)$. It is important to emphasize that the operator \eqref{photon:operator} can be used to describe the \textit{same} photon in two \textit{different} points in spacetime only if spacetime is flat. If the spacetime is curved, then one needs to find a way to account for the effects of propagation. This has already been done \cite{Bruschi:Ralph:2014}, and we report the result below.\footnote{We are using different notation than the one used in the literature \cite{Bruschi:Ralph:2014,Bruschi:Datta:2014,Kohlrus:Bruschi:2017} in order to reduce the burden of the mathematical aspects in favour of clarity of the exposition of the concepts.} 

Let us assume that Alice and Bob are placed at the two different heights $r_{\textrm{A}}$ and $r_{\textrm{B}}$ in the gravitational field of the Earth respectively. Alice is given a photon source and a detector, and prepares a photon which is defined by the function $F_{\omega_0}(\omega)$ that is peaked around $\omega_0$ with width $\sigma$. Bob is given a photon source and a detector identical to those in Alice's possession. Therefore, when Bob receives a photon from Alice he expects to detect a wavepacket characterized by the function $F_{\omega_0}(\omega)$ \textit{as defined locally} by his measuring devices. Nevertheless, the photon that Bob receives is different than the reference photon $F_{\omega_0}(\omega)$ he expects. In particular, it was shown  that he will detect a photon characterized by the frequency distribution $F_{\omega_0'}'(\omega)$, which can be related to the expected one $F_{\omega_0}(\omega)$ through the relation 
\begin{align}\label{photon:operator:bob}
F_{\omega_0'}'(\omega)=\chi(r_{\textrm{A}},r_{\textrm{B}})\,F_{\omega_0}(\chi^2(r_{\textrm{A}},r_{\textrm{B}})\,\omega).
\end{align}
Here, the function $\chi(r_{\textrm{A}},r_{\textrm{B}})$ encodes the overall effect, and we have $\omega\,(h(r)-t)\rightarrow\chi^2(r_{\textrm{A}},r_{\textrm{B}})\omega\,(h(r)-\chi^{-2}(r_{\textrm{A}},r_{\textrm{B}})t)$. In the case of Alice located on the surface of the Earth and Bob orbiting a nonrotating spherical planet, we have that $\chi(r_{\textrm{A}},r_{\textrm{B}})$ reads
\begin{align}\label{chi:expression}
\chi(r_{\textrm{A}},r_{\textrm{B}}):=\sqrt[4]{\frac{1-\frac{3}{2}\frac{r_{\textrm{S}}}{r_{\textrm{B}}}}{1-\frac{r_{\textrm{S}}}{r_{\textrm{A}}}}}.
\end{align}
A pictorial representation can be found in Figure~\ref{fig0}.

The effects of rotation have already been evaluated \cite{Kohlrus:Bruschi:2017}, and can be included to generalize the expression \eqref{chi:expression} of $\chi(r_{\textrm{A}},r_{\textrm{B}})$. However, since we will focus on Earth-based applications where the rotation gives negligible corrections, we keep the given expression \eqref{chi:expression}. An important observation is that, in \eqref{chi:expression}, the factor of $2$ is used for observers on the surface of the Earth, while the factor of $3$ is used for observers orbiting our planet. The expression \eqref{chi:expression} needs to be changed accordingly if the users exchange roles, or assume different states of motion.

Finally, we assume that the phase $\exp[-i\,\omega\,(h(r)-t)]$ induced in \eqref{photon:operator} by the propagation through (weakly) curved spacetime can be compensated by a precise phase reference such as a narrow band reference laser or a radio frequency link~\cite{Gregory2012}.  
For this reason, we drop this phase throughout our work.

%---------------------------------------------------------------------------------------------------%
\section{Distinguishing between wavepacket translation and genuine distortion due to gravity}\label{section:main}
%---------------------------------------------------------------------------------------------------%
This section contains our main proposal. A full justification will be presented in detail, while explicit calculations can be found in the relevant appendix.

%---------------------------------------------------------------------------------------------------%
\subsection{Gravity effects on quantum coherence}
%---------------------------------------------------------------------------------------------------%
As described above there is a relation between the wavepacket $F_{\omega_0}(\omega)$ that Bob expects and the wavepacket $F_{\omega_0'}'(\omega)$ that he will receive from Alice. A crucial aspect that we want to study is the potential difference of the effects for coherent photons (i.e., photons characterized by pure quantum states) and incoherent photons (those characterized by completely mixed quantum states). The reason is that, while pure states $\hat{\rho}_{\textrm{p}}$ and mixed states $\hat{\rho}_{\textrm{m}}$ may have the same contribution for each frequency $\omega$ within the frequency distribution $F_{\omega_0}(\omega)$, i.e., the diagonal elements of the states are the same, in the mixed state case the phase relation between the different frequency components is not fixed. Since the study of the effects of gravity on genuine quantum features, such as quantum coherence, is a topic of great interest, we will apply our methods to both classes of states and compare the outcomes.

To simplify our analysis we introduce $|1_{\omega}\rangle:=\hat{a}^\dag_\omega|0\rangle$, $|1^{\textrm{F}_{\omega_0}}\rangle:=\int d\omega F_{\omega_0}(\omega) |1_{\omega}\rangle$ and $\hat{\rho}_{\textrm{p}}(0):=|1^{\textrm{F}_{\omega_0}}\rangle\langle1^{\textrm{F}_{\omega_0}}|$. We note that purely diagonal mixed states of sharp continuous frequencies are ill defined (see Appendix \ref{appendix:one}). Therefore, we introduce the ``rectangular'' single-photon states $|1^{(n)}(\theta)\rangle$ with finite width $\sigma_*$ defined by
\begin{align}\label{window:state}
|1^{(n)}(\theta)\rangle:=&\int_{(n-1/2)\sigma_*}^{(n+1/2)\sigma_*}\frac{d\omega}{\sqrt{\sigma_*}}e^{i\,\theta\,\frac{\omega}{\sigma_*}}|1_{\omega}\rangle.
\end{align}
Crucially, we assume that $\lambda:=\sigma_*/\sigma\ll1$, that is, the width of these rectangular states is much smaller than that of the frequency distribution that defines our system (that is, that defines the width of the wavepacket). It is easy to check that $\langle1^{(n)}(\theta)|1^{(m)}(\theta)\rangle=\delta_{nm}$.

We introduce on the photon states
\begin{align}\label{photon:states}
\hat{\rho}_{\textrm{m}}(0):=&\sum_n \lambda\,|F_{\omega_0}(\chi^2\,\lambda\,n)|^2\, |1^{(n)}(0)\rangle\langle1^{(n)}(0)|\nonumber\\
\hat{\rho}_{\textrm{p}}(0)=&\int d\omega d\omega' F_{\omega_0}(\omega)F^*_{\omega_0}(\omega') |1_{\omega}\rangle\langle1_{\omega'}|,
\end{align}
which will underpin all of the following analysis.

The definitions \eqref{photon:states} give states that are correctly normalized as $\textrm{Tr}(\hat{\rho}_{\textrm{m}}(0))=\textrm{Tr}(\hat{\rho}_{\textrm{p}}(0))=1$. Finally, as mentioned above, we have that $\langle1^{(n)}(0)|\hat{\rho}_{\textrm{m}}(0)|1^{(n)}(0)\rangle=\langle1^{(n)}(0)|\hat{\rho}_{\textrm{p}}(0)|1^{(n)}(0)\rangle\approx|F_{\omega_0}(\chi^2\,\lambda\,n)|^2$, which implies that local measurements on the single rectangle photon state $|1^{(n)}(0)\rangle$ cannot distinguish between the two states. This emphasizes the role of the off-diagonal terms in the density matrices, which describe for the quantum coherence between Fock states.

The received states $\hat{\rho}_{\textrm{m}}(\chi)$ and $\hat{\rho}_{\textrm{p}}(\chi)$ by Bob read
\begin{align}\label{photon:states:modified}
\hat{\rho}_{\textrm{m}}(\chi)=&\chi^2\sum_n \lambda\,|F_{\omega_0}(\chi^2\,\lambda\,n)|^2\, |1^{(n)}(0)\rangle\langle1^{(n)}(0)|\nonumber\\
\hat{\rho}_{\textrm{p}}(\chi)=&\chi^2\int d\omega d\omega' F_{\omega_0}(\chi^2\omega)F^*_{\omega_0}(\chi^2\omega') |1_{\omega}\rangle\langle1_{\omega'}|.
\end{align}

%---------------------------------------------------------------------------------------------------%
\subsection{Change of quantum coherence due to gravity}
%---------------------------------------------------------------------------------------------------%
We can now ask ourselves a preliminary question: \textit{does the coherence of the state change due to gravity?} To answer this question we use the purity $\gamma$ of a quantum state $\hat{\rho}$ defined as $\gamma(\hat{\rho}):=\textrm{Tr}[\hat{\rho}^2]$, which is bounded by $0\leq\gamma(\hat{\rho})\leq1$ and is equal to unity only for pure states.\footnote{Contrary to finite $d$-dimensional systems, for which $\frac{1}{d}\leq\gamma\leq1$.}
Introducing the purities $\gamma_{\textrm{m}}(\chi):=\textrm{Tr}[\hat{\rho}_{\textrm{m}}^2(\chi)]$ and $\gamma_{\textrm{p}}(\chi):=\textrm{Tr}[\hat{\rho}_{\textrm{p}}^2(\chi)]$, we show in Appendix~\ref{appendix:one} that $\gamma_{\textrm{m}}(0):=\gamma(\hat{\rho}_{\textrm{m}}(0))=\int d\omega |F_{\omega_0}(\omega)|^4$ and $\gamma_{\textrm{p}}(0):=\gamma(\hat{\rho}_{\textrm{p}}(0))=1$. Using the modified states \eqref{photon:states:modified} \textit{in Bob's local frame} we find that
\begin{align}\label{photon:states:modified:coherence}
\gamma_{\textrm{m}}(\chi)=&\gamma_{\textrm{m}}(0)\nonumber\\
\gamma_{\textrm{p}}(\chi)=&\gamma_{\textrm{p}}(0).
\end{align}
This means that, although the wavepackets have been affected, the mixedness of the quantum states does not change and there is no interaction of the system with any additional degrees of freedom (such as gravity). This is consistent with the semiclassical approach provided by quantum field theory in curved spacetime.

%---------------------------------------------------------------------------------------------------%
\subsection{Quantifying genuine distortion due to gravity}
%---------------------------------------------------------------------------------------------------%
We are now able to move on to the main proposal. The key observation is that many protocols of interest to Bob will \textit{depend on how well the na\"ively expected state $\hat{\rho}$ (or wavepacket $F_{\omega_0}$) and received state $\hat{\rho}'$ (or wavepacket $F'_{\omega_0'}$) overlap}. 
We are thus motivated to introduce the overlap $\Delta:=\sqrt{\mathcal{F}(\hat{\rho}',\hat{\rho})}$ between the states (wavepackets) of the two systems using the quantum fidelity $\mathcal{F}(\hat{\rho}',\hat{\rho}):=\textrm{Tr}(\sqrt{\sqrt{\hat{\rho}}\hat{\rho}'\sqrt{\hat{\rho}}})$, following the results of the original work in this direction \cite{Bruschi:Ralph:2014}. It is then convenient to write  $F_{\omega_0}(\omega)=f_{\omega_0}(\omega)\exp[i\,\psi(\omega)]$, where $f_{\omega_0}(\omega):=|F_{\omega_0}(\omega)|$ is the modulus of the frequency distribution and $\psi(\omega)$ is its phase. Using this decomposition, in Appendix~\ref{appendix:one} we show that the single-photon overlap $\Delta$ for the two choices of initial single-photon states reads
\begin{align}\label{photon:overlap}
\Delta_{\textrm{p}}=&\chi\left|\int_{-\infty}^{+\infty}d\omega f_{\omega_0}(\chi^2 \omega)f_{\omega_0}(\omega)\,e^{i(\psi(\chi^2 \omega)-\psi(\omega))}\right|\nonumber\\
\Delta_{\textrm{m}}=&\chi\int_{-\infty}^{+\infty}d\omega f_{\omega_0}(\chi^2\,\omega)\,f_{\omega_0}(\omega)
\end{align}
for the mixed and pure state scenarios respectively. Here we have defined $\chi:=\chi(r_{\textrm{A}},r_{\textrm{B}})$ for convenience of presentation. Note that we have extended the lower limit of integration $-\infty$ for algebraic purposes, a modification that requires us to assume that all peaked functions $F$ considered, together with any transformed ones $F'$ obtained from such functions, always remain far from the origin of the frequency axis. Importantly, we know that $0\leq\Delta_{\textrm{p}}\leq\Delta_{\textrm{m}}\leq1$ due to basic properties of integrals, which suggests already that the phase $\psi(\omega)$ present in the pure state can only \textit{increase} the effects of gravity, that is, reduce the overlap $\Delta$.
We emphasize that the quantities $\Delta$ \textit{fully quantify} the effects of gravity induced on the photons sent by Alice and received by Bob \cite{Bruschi:Ralph:2014}. 

We are interested at this point in separating two effects that occur simultaneously: the \textit{rigid translation} of the wavepacket and its \textit{genuine deformation}. 
To define concretely the two effects we first note that we expect any physical system, to be used for realistic quantum information tasks, to be localized in space and time. In practical terms, we can model such a system through a peaked function $f_{\omega_0}(\omega)$ that depends functionally on $\omega$ as $f_{\omega_0}(\omega)\equiv 1/\sqrt{\sigma}\,\tilde{f}(\frac{\omega-{\omega_0}}{\sigma})$. This allows us to introduce a dimensionless function $\tilde{f}$. Furthermore, the phase $\psi(\omega)$ of the overall wavepacket is functionally dependent on $\omega$ through $\psi(\omega/\sigma)$, since $\omega$ is a dimensionful quantity. Using the expression for single-photon states, we have
\begin{align}\label{photon:overlap:second}
\Delta_{\textrm{p}}=&\chi\left|\int_{-\infty}^{+\infty}d\omega \tilde{f}\left(\frac{\chi^2 \omega-\omega_0}{\sigma}\right) \tilde{f}\left(\frac{\omega-{\omega_0}}{\sigma}\right) e^{i\delta\psi(\omega)}\right|\nonumber\\
\Delta_{\textrm{m}}=&\chi\int_{-\infty}^{+\infty}d\omega \tilde{f}\left(\frac{\chi^2 \omega-\omega_0}{\sigma}\right) \tilde{f}\left(\frac{\omega-{\omega_0}}{\sigma}\right)
\end{align}
where $\delta\psi(\omega):=\psi(\chi^2 \frac{\omega}{\sigma})-\psi(\frac{\omega}{\sigma})$ here only for convenience of presentation.

\begin{figure}[ht!]
    \centering
    \includegraphics[width=0.8\linewidth]{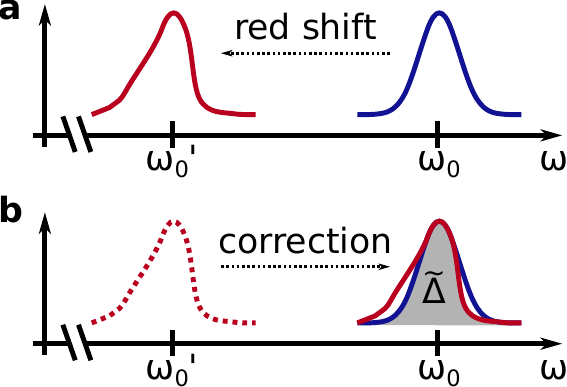}
    \caption{\textit{Simple solution to the gravitationally-induced wavepacket separation.} Panel (a) depicts the gravitational red-shift, while panel (b) illustrates a simple correction obtained by rigid translation of the whole wavepacket. In  (a), the central frequency is shifted from $\omega_0$ to $\omega_0'$, and the shape distribution is distorted. In (b), a correction is applied: The \textit{whole} new wavepacket (dotted) is shifted back by the \textit{same} amount that the central frequency $\omega_0$ has changed. Due to the distortion, the overlap to the initial  wavepacket is reduced to the overlap $\tilde{\Delta}$, emphasized in gray in the figure.}
    \label{fig1}
\end{figure}

It is clear that, if we were to set $\chi=1$ in the term $\tilde{f}(\frac{\chi^2 \omega-\omega_0}{\sigma})$, the peaks of the two distributions would coincide. Therefore, we first write 
\begin{align*}
\chi^2 \omega-\omega_0=\chi^2 (\omega-\omega_0+\omega_0)-\omega_0=\chi^2 [(\omega-\omega_0)+\kappa\omega_0]
\end{align*} 
with $\kappa:=(\chi^2-1)/\chi^2$. We then view it natural to define the quantity $\kappa\omega_0$ as the \textit{rigid shift} or \textit{classical gravitational redshift} of the incoming wavepacket, which Bob might want to compensate for. At this point, the temptation would be to define the genuine deformation by the quantities \eqref{photon:overlap:second} obtained as the overlap between the incoming \textit{corrected} wavepacket and the expected wavepacket. In other words, by implementing the replacement $\omega\rightarrow\omega-\kappa\,\omega_0$ in the first function (and phase) of the overlaps \eqref{photon:overlap:second}. A pictorial representation of this approach can be found in Figure~\ref{fig1}, where we have introduced the corrected overlap $\tilde{\Delta}$ (defined properly below).

This correction procedure, however, is na\"ive  since it could in principle lead to underestimation of the true genuine distortion. The reason is that, in general, the affected wavepacket will not be distorted in a way to fit completely within, or cover completely, the original one (i.e., as we have depicted in panel (b) of Figure~\ref{fig1}) after the correction described above has been applied. One part, e.g., on the right, might fall under the original wavepacket, while some area on the left might be left outside, which leaves open the question of how to optimize between these two areas. A more refined correction procedure is therefore necessary.

To achieve this goal, we assume that Bob can indeed \textit{rigidly shift} the entire wavepacket of the \textit{received photon} by a finite constant amount $\delta\omega$, thereby translating the whole frequency spectrum in a chosen direction in the frequency domain. Concretely, we assume that he is able to perform the transformation $\omega\rightarrow\omega+\delta\omega$ \textit{for all $\omega$} (once more, if he performs the rigid shift $\delta\omega=-\kappa\omega_0$ he will not obtain, in general, the optimal overlap. See panel (b) of Figure~\ref{fig1}). 

\begin{figure}[ht!]
    \centering
    \includegraphics[width=0.8\linewidth]{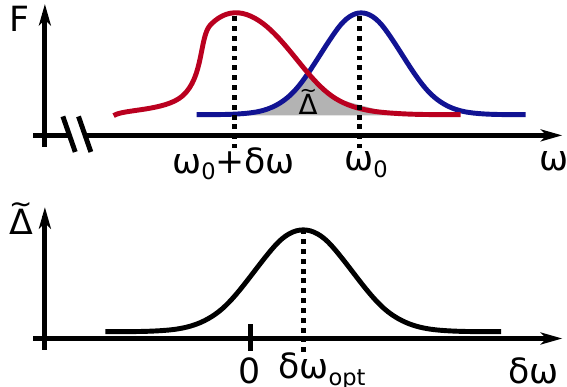}
    \caption{\textit{Quantification of the genuine distortion}. After the correction shown in Figure~\ref{fig1}, the corrected overlap $\tilde{\Delta}$ of the wavepackets is not guaranteed to be optimal. Therefore, Bob chooses instead to \textit{rigidly shift} the \text{whole} incoming wavepacket by the freely adjustable amount $\delta \omega$. This provides an overlap $\tilde{\Delta}$ as a function of $\delta\omega$ which can be maximized. The lower panel shows a plot of $\tilde{\Delta}$ versus $\delta \omega$ with the optimal overlap $\delta \omega_{\textrm{opt}}$ marked.}
    \label{fig2}
\end{figure}

However, and this is the key observation, instead of choosing to shift the wavepacket by the amount $\delta\omega=-\kappa\omega_0$ that we found above, Bob chooses to seek the shift $\delta\omega_{\textrm{opt}}$ that results in the largest value $\Delta_{\textrm{opt}}$ of the overlap \eqref{photon:overlap}. This will simultaneously achieve two goals: (i) it will define the \textit{effective rigid shift} of the wavepacket, and (ii) it will provide us with a value $\Delta_{\textrm{opt}}$ that quantifies the \textit{genuine distortion} of the wavepacket due to gravity. The optimization procedure is depicted in Figure~\ref{fig2}.

The above reasoning, therefore, motivates us to define the new overlap quantities $\tilde{\Delta}$ as
\begin{align}\label{photon:overlap:new}
\tilde{\Delta}_{\textrm{p}}=&\left|\int_{-\infty}^{+\infty}dz \tilde{f}(\chi z+\bar{z})\tilde{f}(z/\chi)e^{i\,\delta\psi(z)}\right|\nonumber\\
\tilde{\Delta}_{\textrm{m}}=&\int_{-\infty}^{+\infty}dz \tilde{f}(\chi z+\bar{z})\tilde{f}(z/\chi),
\end{align}
which are corrected overlaps where we have introduced the quantities $z:=\chi\frac{\omega-\omega_0}{\sigma}$, $z_0:=\frac{\omega_0}{\sigma}$, $\delta z:=\frac{\delta\omega}{\sigma}$ and $\bar{z}:=\chi^2[\kappa z_0+\delta z]$. The variable $\bar{z}$ is nothing more than the conveniently rescaled and shifted $\delta\omega$. We have also introduced $\delta\psi(z):=\psi(\chi z+z_0+\bar{z})-\psi(z/\chi+z_0)$ here only for convenience of presentation.

Finding the stationary points of the overlaps \eqref{photon:overlap:new} with respect to $\bar{z}$ (the conveniently relabelled $\delta\omega$) and then using the optimal value $\bar{z}_{\textrm{opt}}$ to evaluate the maximum $\tilde{\Delta}_{\textrm{opt}}$ finally quantifies the genuine distortion due to gravity. Furthermore, the value $\bar{z}_{\textrm{opt}}$ also provides the value $\delta\omega_{\textrm{opt}}$ that quantifies the effective rigid shift of the whole wavepacket. The comparison between the results for the pure and mixed state case then quantitatively informs us on the difference of the effects of gravity on the quantum coherence of the photonic states.  
This is our main result.

%---------------------------------------------------------------------------------------------------%
\subsection{Optimal overlap of wavepackets with a well-defined global peak}
%---------------------------------------------------------------------------------------------------%
We have defined our main quantity \eqref{photon:overlap:new} as an optimized comparison between peaked photonic wavepackets. We note here that this might not be a well defined operation for all possible shapes of wavepacket. For example, if $\tilde{f}$ contains two distinct and comparable peaks, it remains unclear that the procedure outlined before would, in the end, provide one optimal overlap. It is reasonable, however, to expect that our optimization protocol will provide a meaningful answer if the wavepacket contains a well-defined shape with a global maximum. By this we mean, for example, that the function $\tilde{f}$ is obtained by a product of two functions $\tilde{f}_{\textrm{env}}$ and $\tilde{f}_{\textrm{ps}}$, where $\tilde{f}_{\textrm{env}}$ provides a peaked envelope shape to the function, while the $\tilde{f}_{\textrm{ps}}$ provides some particular structure to $\tilde{f}$. An example is $\tilde{f}=C\,\exp[-(z-z_0)^2]\cos^2[z]$. 

In this work we will focus on functions $\tilde{f}$ that have a Gaussian envelope peaked around a given frequency $\omega_0$. As we will see below when considering specific examples, the overlaps $\tilde{\Delta}$ will be optimized for effectively $\bar{z}=0$, which in turn means that the optimal shift $\delta\omega_{\textrm{opt}}$ is (up to negligible corrections in some cases) the rigid shift factor $\kappa\,\omega_0$.

%---------------------------------------------------------------------------------------------------%
\subsection{Wavepackets with multiple photons}
%---------------------------------------------------------------------------------------------------%
Many photonic quantum information protocols, such as quantum key distribution ones, are designed to employ single photons \cite{Pirandola:Andresen:2020,Xu:Ma:2020}. However, many others employ pulses of light instead \cite{Braunstein:vanLoock:2005}. A pulse of light is composed of many photons and we therefore wish to extend our single-photon result \eqref{photon:overlap:new} to different states: Fock states, coherent states and single-mode squeezed states.

To do so we observe the fact that the quantity $\tilde{\Delta}$ in \eqref{photon:overlap} has been obtained as the overlap of two different single-photons states: the expected one and the effectively received one by Bob. We can therefore extend the concept and define the quantity $\tilde{\Delta}^{(N)}$ that measures the genuine distortion due to gravity of a wavepacket composed of an average of $N$ photons. Importantly, we set the number $N$ of photons as the figure of merit that will allow us to compare the scaling of the results. Since we have already defined $\Delta:=\sqrt{\mathcal{F}(\hat{\rho}',\hat{\rho})}$ above, in cases such as those where we want to make explicit the average number $N$ of photons we choose to emphasize that this number in the state as $\hat{\rho}^{(N)}$, and therefore write $\Delta^{(N)}:=\sqrt{\mathcal{F}(\hat{\rho}^{(N)'},\hat{\rho}^{(N)})}$.
We also to introduce the useful quantity
\begin{align}\label{complex:overlap}
\tilde{\Lambda}_{\textrm{p}}=&\int_{-\infty}^{+\infty}dz \tilde{f}(\chi z+\bar{z})\tilde{f}(z/\chi)e^{i\,\delta\psi(z)}
\end{align}
and note that $\tilde{\Delta}^{(1)}_{\textrm{p}}\equiv|\tilde{\Lambda}_{\textrm{p}}|$.

%---------------------------------------------------------------------------------------------------%
\subsubsection{$N$-photon Fock state}
%---------------------------------------------------------------------------------------------------%
We start with the initial $N$-photon Fock state $\hat{\rho}_{\textrm{p}}^{(N)}(0):=|N_{\omega_0}\rangle\langle N_{\omega_0}|$, which is the generalization to $N$ photons of the state $\hat{\rho}_{\textrm{p}}(0)$ in \eqref{photon:states}. In this case, we can compute the overlap $\tilde{\Delta}^{(N)}$, and find $\tilde{\Delta}^{(N)}=\tilde{\Delta}^N$, which implies that the fidelity of the $N$-photon Fock state is polynomial in the number of photons. Note that, in this case, we do not have a mixed-state counterpart to the ($N$-photon) Fock state.

%---------------------------------------------------------------------------------------------------%
\subsubsection{Coherent states}
%---------------------------------------------------------------------------------------------------%
We now focus on two different populated states for our system: a coherent state $\hat{\rho}_{\textrm{p}}^{(N)}(0):=|\alpha\rangle\langle\alpha|$, defined through $|\alpha\rangle:=\exp[\alpha\,\hat{a}^\dag_{\omega_0}-\alpha^*\,\hat{a}_{\omega_0}]|0\rangle$ with average number $N=|\alpha|^2$ of photons, and its ``mixed counterpart'' $\hat{\rho}_{\textrm{m},N}(0):=\exp[-|\alpha|^2]\sum_{n=0}^{\infty}|\alpha|^{2n}/n!|n\rangle\langle n|$, obtained by adding a phase $e^{i\theta}$ to $\alpha$ and integrating over $\theta$ (see \cite{vanEnk:Fuchs:2001,VanEnk:Fuchs:2002}).
We know that a coherent state can be used, for example, to model the state of the light emitted by a laser \cite{vanEnk:Fuchs:2001,VanEnk:Fuchs:2002}, which is paramount in many applications of continuous variables quantum information processing~\cite{Braunstein:vanLoock:2005}.

Simple algebra allows us to obtain
\begin{align}\label{photon:overlap:coherent:states}
\tilde{\Delta}^{(N)}_{\textrm{p}}=&e^{-(1-\Re\tilde{\Lambda}_{\textrm{p}})\,N}\nonumber\\
\tilde{\Delta}^{(N)}_{\textrm{m}}=&\tilde{\Delta}^{(1)}_{\textrm{m}}.
\end{align}
This result shows that the coherent state, which by definition contains quantum coherence of different Fock states of the photonic Hilbert space, enjoys an enhanced effect (i.e., smaller overlap) as compared to the mixed counterpart. In particular, if it is true that there is an effect for one photon, and therefore $\Re\tilde{\Lambda}_{\textrm{p}}<1$, this implies that there is an exponential gain by considering such coherent states. In those scenarios, as in our cases considered below, one has $\Re\tilde{\Lambda}_{\textrm{p}}\approx 1- \delta_1$, with $\delta_1\ll1$, and therefore we have that $\Re\tilde{\Lambda}_{\textrm{p}}\approx e^{- \delta_1\,N}\approx(1- \delta_1\,N)$. This can be of aid when designing concrete experimental applications that employ the predictions of this work.

%---------------------------------------------------------------------------------------------------%
\subsubsection{Single-mode squeezed states}
%---------------------------------------------------------------------------------------------------%
We finally focus on single-mode squeezed states. Single-mode squeezed states are nonclassical states \cite{Schnabel:2017} that can be produced, for example, using stimulate emission via nonlinear crystals, a process known as degenerate parametric down-conversion \cite{Klauder:1986,Andersen:Gehring:2016}. In this case, we study the single-mode squeezed state $\hat{\rho}_{\textrm{p}}^{(N)}(0):=|s\rangle\langle s|$ defined through $|s\rangle:=\sum_{n=0}^{+\infty}\tanh^n (s)/\cosh (s)|2n\rangle$ with average number $N=2\,\sinh^2(s)$ of photons, and its ``mixed counterpart'' $\hat{\rho}_{\textrm{m},N}(0):=\sum_{n=0}^{+\infty}\tanh^{2n} (s)/\cosh^2 (s)|2n\rangle\langle 2n|$. Here $s$ is called the squeezing parameter.

The same algebra used above gives us
\begin{align}\label{photon:overlap:squeezed:states}
\tilde{\Delta}^{(N)}_{\textrm{p}}=&\left(\left(1+\frac{1}{2}\left(1-\Re\tilde{\Lambda}_{\textrm{p}}\right) N\right)^2+\frac{1}{4} (\Im\tilde{\Lambda}_{\textrm{p}})^2 N^2\right)^{-1/2}\nonumber\\
\tilde{\Delta}^{(N)}_{\textrm{m}}=&\tilde{\Delta}^{(1)}_{\textrm{m}}.
\end{align}
Note that if $\tilde{\Lambda}_{\textrm{p}}\in\mathbb{R}$, then we have the simplified expression $\tilde{\Delta}^{(N)}_{\textrm{p}}=\bigl(1+(1-\tilde{\Delta}^{(1)}_{\textrm{p}})N/2\bigr)^{-1}$. Finally, for $N\,\tilde{\Delta}^{(1)}_{\textrm{p}}\gg1$ we have $\tilde{\Delta}^{(N)}_{\textrm{p}}\approx2/(N\,\tilde{\Delta}^{(1)}_{\textrm{p}})$.

%---------------------------------------------------------------------------------------------------%
\subsection{Extension to wavepackets with multiple peaks}
%---------------------------------------------------------------------------------------------------%
We note that not all wavepackets will have the simple property $f_{\omega_0}(\omega)\propto \tilde{f}(\frac{\omega-{\omega_0}}{\sigma})$, where $\omega_0$ identifies the only peak of the distribution. Here we consider the more general case of frequency distributions with a clear global maximum, but also other local maxima. This will prove especially useful when considering specific applications, where multiple peaks are present.

Let the function $f_{\omega_0}(\omega)$ have the expression 
\begin{align}
f_{\omega_0}(\omega)\equiv \frac{e^{i\psi(\omega)}}{\sqrt{\sigma}} \tilde{f}\left(\frac{\omega-{\omega_0}}{\sigma}\right)\sum_n \tilde{G}_n\left(\frac{\omega-\omega_{0,n}}{\mu_n}\right),
\end{align}
where the functions $\tilde{G}_n$ are positive and are peaked around $\omega_{0,n}$ with width  $\mu_n$. Here, the index $n$ runs over (a subset of) the integer numbers $\mathbb{Z}$. Recall that the original functions $f_{\omega_0}(\omega)$ are normalized such that $\int_0^{\infty}d\omega|f_{\omega_0}(\omega)|^2=1$. Then, we repeat the analysis above and obtain the generalized version of \eqref{photon:overlap:new}, which reads
\begin{align}\label{photon:overlap:more:peakes}
\tilde{\Delta}_{\textrm{p}}=&\left|\sum_{n,m}\int_{-\infty}^{+\infty}dz \tilde{f}\left(\chi\,z+\bar{z}\right) \tilde{f}(z/\chi) e^{i\,\delta\psi(z)}\right.\nonumber\\
&\times\tilde{G}_n\left(\frac{\sigma}{\mu_n}\left(\chi\,z+z_0+\bar{z}-\omega_{0,n}/\sigma\right)\right)\nonumber\\
&\left.\times\tilde{G}_m\left(\frac{\sigma}{\mu_m}\left(z/\chi+z_0-\omega_{0,n}/\sigma\right)\right)\right|
\end{align}
in the pure case, and
\begin{align}
\tilde{\Delta}_{\textrm{m}}=&\sum_{n}\int_{-\infty}^{+\infty}dz \tilde{f}\left(\chi\,z+\bar{z}\right) \tilde{f}(z/\chi) \nonumber\\
&\times\tilde{G}_n\left(\frac{\sigma}{\mu_n}\left(\chi\,z+z_0+\bar{z}-\omega_{0,n}/\sigma\right)\right)\nonumber\\
&\times\tilde{G}_m\left(\frac{\sigma}{\mu_n}\left(z/\chi+z_0-\omega_{0,n}/\sigma\right)\right)
\end{align}
in the mixed one.
Here we have introduced $\delta\psi(z):=\psi(\chi z+z_0+\bar{z})-\psi(z/\chi+z_0)$ for this case.
This expression reduces to \eqref{photon:overlap:new} in different cases, such as when $\sum_n \tilde{G}_n=1$. We can also include the complex part of the $G$-functions, but this is an unnecessary complication at this point and we leave it to future work.

%---------------------------------------------------------------------------------------------------%
\section{Applications}\label{section:applications}
%---------------------------------------------------------------------------------------------------%
In this section we discuss applications of our results.
As explained above, we will focus on frequency profiles that have a clearly identifiable global peak. Note that all functions $F(z)$ are normalized such that $\int dz |F(z)|^2=\int dz f^2(z)=1$. This is equivalent to say that the quantum states of the electromagnetic field have unit norm.

%---------------------------------------------------------------------------------------------------%
\subsection{Near-Earth protocols}
%---------------------------------------------------------------------------------------------------%
Our scheme can be applied, in principle, to arbitrary spacetimes where a notion of gravitational redshift can be meaningfully introduced. Here, however, we focus our methods to the spacetime surrounding the Earth, which can be modelled as described above. 
In the near-Earth regime we have $r_{\textrm{S}}/r_{\textrm{A}}\ll1$ and $r_{\textrm{S}}/r_{\textrm{B}}\ll1$, and all the perturbative analysis performed in the following is governed by parameters which are of the order of $\sim 10^{-9}$, or smaller. Therefore, we introduce the parameter $\delta=\delta_1+\delta_2\ll1$ through 
\begin{align}
\chi=1+\delta=1+\delta_1+\delta_2,
\end{align}
where $\delta_1$ and $\delta_2\sim\delta_1^2\ll\delta_1$ collect the first and second order corrections of $\delta$ respectively, as we explain below. Note that $\delta_1\gg\delta_2$ as long as $r_{\textrm{B}}$ is far away from $3/2r_{\textrm{A}}$ (given the particular choice of the expression $\chi$). We obtain 
\begin{align}\label{approximate:delta}
\delta\approx \underbrace{\frac{1}{4}\frac{r_{\textrm{S}}}{r_{\textrm{A}}}-\frac{3}{8}\frac{r_{\textrm{S}}}{r_{\textrm{B}}}}_{\delta_1}+\underbrace{\frac{5}{32}\frac{r_{\textrm{S}}^2}{r_{\textrm{A}}^2}-\frac{3}{32}\frac{r_{\textrm{S}}^2}{r_{\textrm{A}}r_{\textrm{B}}}-\frac{27}{128}\frac{r_{\textrm{S}}^2}{r_{\textrm{B}}^2}}_{\delta_2}.
\end{align}
In case the location of Bob is far enough from Alice (and, therefore, the Earth), it follows that his own contributions to \eqref{approximate:delta} are negligible. This implies that \eqref{approximate:delta} reads $\delta\approx \frac{1}{2}\frac{r_{\textrm{S}}}{r_{\textrm{A}}}+\frac{5}{32}\frac{r_{\textrm{S}}^2}{r_{\textrm{A}}^2}$.

Finally, assuming that Bob is located at a higher altitude than Alice, and that the distance $L:=r_{\textrm{B}}-r_{\textrm{A}}$ between the two is (much) smaller than $r_{\textrm{A}}$, i.e., if $L/r_{\textrm{A}}\ll 1$, then \eqref{approximate:delta} reduces to 
\begin{align}\label{approximate:delta:two}
\delta\approx \underbrace{-\frac{1}{8}\frac{r_{\textrm{S}}}{r_{\textrm{A}}}}_{\delta_1}+\underbrace{\frac{3}{8}\frac{r_{\textrm{S}}\,L}{r_{\textrm{A}}^2}-\frac{19}{128}\frac{r_{\textrm{S}}^2}{r_{\textrm{A}}^2}}_{\delta_2}.
\end{align}
These expressions, in particular \eqref{approximate:delta}, will be used below to simplify the overlaps that we have computed before, considering realistic space-based  scenarios.

%---------------------------------------------------------------------------------------------------%
\subsection{Application to bell-shaped wavepackets}
%---------------------------------------------------------------------------------------------------%
We now proceed by considering two different wavepackets with an overall Gaussian envelope: (i) a simple Gaussian wavepacket
and (ii) a frequency-comb wavepacket modulated by a Gaussian. We have chosen these wavepackets, since Gaussian frequency distributions can be seen as a crude approximation to what is usually employed for pulse generation or filtering. These realistic profiles include those obtained as the impulse response of a root-raised-cosine filter, and are related to sinc-like functions \cite{Daumont:Rihawi:2008}. A more realistic modelling, which might require numerical analysis, is beyond the scope of this work. Recall that, in the following, we give the functions in terms of the normalized and rescaled variable  $z:=\chi\frac{\omega-\omega_0}{\sigma}$. Furthermore, we only consider overlap functions for single photon wavepackets. Note that the phases $\phi$ used below have nothing to do with the angular coordinate in the metric. Extensions to multiple photons have been discussed above. Finally, all explicit computations can be found in Appendix \ref{appendix:application:to:concrete:wavepackets}.

%---------------------------------------------------------------------------------------------------%
\subsubsection{Gaussian wavepackets}
%---------------------------------------------------------------------------------------------------%
We start by considering a Gaussian frequency profile. It is known that Gaussian distributions are resilient to modifications of the type discussed in this work \cite{Kish:Ralph:2016}. For this reason, we can already anticipate that the choice of such functions \textit{without any complex phase} will result into small deformations. We will consider two Gaussians with different phase factors, that is, two different functions $\psi(\omega)$. We will show that the choice of functional dependence of the phase $\psi(\omega)$ on the frequency $\omega$ dramatically changes the results.

\textit{Linear phase.} The first choice of wavepacket is the normalized Gaussian $F(z)=1/\sqrt[4]{2\,\pi}\exp[-z^2/4-i\,\tilde{\phi}\,z]$. In this case, the phase $\psi(\omega)$ is linear in the frequency and reads $\psi(\omega)=\phi\,\omega$, where $\phi:=\tilde{\phi}/\sigma$ is a constant. This allows us to easily compute the overlaps $\tilde{\Delta}^{\textrm{Ga}}$ as
\begin{align}\label{photon:overlap:Gaussian}
\tilde{\Delta}^{\textrm{Ga}}_{\textrm{p}}=&\frac{\sqrt{2}\,\chi}{\sqrt{1+\chi^4}}\,e^{-\frac{(\chi^2-1)^2}{\chi^4+1}\tilde{\phi}^2}e^{-\frac{\bar{z}}{\chi^4+1}}\nonumber\\
\tilde{\Delta}^{\textrm{Ga}}_{\textrm{m}}=&\frac{\sqrt{2}\,\chi}{\sqrt{1+\chi^4}}e^{-\frac{\bar{z}}{\chi^4+1}}.
\end{align}
As argued above, we now need to maximize the expressions \eqref{photon:overlap:Gaussian} with respect to $\bar{z}$. It is immediate to see that the maximum occurs for $z=\bar{z}_{\textrm{opt}}=0$, which implies that the redshift is indeed quantified by $\delta \omega=-\kappa \omega_0$ and the optimal overlaps are
\begin{align}\label{optimal:photon:overlap:Gaussian}
\tilde{\Delta}^{\textrm{Ga}}_{\textrm{p,opt}}=&\frac{\sqrt{2}\,\chi}{\sqrt{1+\chi^4}}\,e^{-\frac{(\chi^2-1)^2}{\chi^4+1}\tilde{\phi}^2}\nonumber\\
\tilde{\Delta}^{\textrm{Ga}}_{\textrm{m,opt}}=&\frac{\sqrt{2}\,\chi}{\sqrt{1+\chi^4}}.
\end{align}
When we consider the near-Earth regime, we then have
\begin{align}\label{photon:overlap:Gaussian:approximate}
\tilde{\Delta}^{\textrm{Ga}}_{\textrm{p,opt}}\approx&1-(1+2\tilde{\phi}^2)\,\delta_1^2\nonumber\\
\tilde{\Delta}^{\textrm{Ga}}_{\textrm{m,opt}}\approx&1-\delta_1^2,
\end{align}
and we must have $\tilde{\phi}\delta_1\ll1$ for the perturbative approach to be valid. In this case, $\delta \omega\approx-(1-2\delta_1)\omega_0$. 

\textit{Quadratic phase.} We can choose a different phase profile $\psi(\omega)$ for our Gaussian, such as a \textit{quadratic} one $\psi(\omega)=\phi^2\,\omega^2$, with $\phi:=\tilde{\phi}/\sigma$ a constant. Therefore, we employ the normalized Gaussian $F(z)=1/\sqrt[4]{2\,\pi}\exp[-z^2/4-i\,\tilde{\phi}^2\,z^2]$ with our chosen phase profile, and we find
\begin{align}\label{photon:overlap:Gaussian:second:profile}
\tilde{\Delta}^{\textrm{Ga}}_{\textrm{p}}=&\frac{\sqrt{2}\,\chi}{\sqrt{1+\chi^4}}\,\frac{e^{-4\,\frac{(\chi^2-1)^2}{\chi^4+1}\frac{\tilde{\phi}^4}{\xi(\tilde{\phi})}z_0^2}}{\sqrt{\xi(\tilde{\phi})}}e^{-16\,a_1(\tilde{\phi}^2) \bar{z}-\frac{1}{4}\,a_2(\tilde{\phi}^2) \bar{z}^2}\nonumber\\
\tilde{\Delta}^{\textrm{Ga}}_{\textrm{m}}=&\frac{\sqrt{2}\,\chi}{\sqrt{1+\chi^4}}e^{-\frac{1}{4}\,a_2(0) \bar{z}^2},
\end{align}
where we have introduced $\xi(\tilde{\phi}):=1+16\tilde{\phi}^4\frac{(\chi^4-1)^2}{(\chi^4+1)^2}$, $a_1(\tilde{\phi})=\chi^2\frac{(\chi^2-1)^2}{(\chi^4+1)^2}\frac{\tilde{\phi}^2}{\xi}z_0$ and $a_2(\tilde{\phi}):=\frac{1+16\tilde{\phi}^4}{\chi^4+1}\frac{1}{\xi}$ as $\tilde{\phi}$-dependent functions.
Optimizing the expressions \eqref{photon:overlap:Gaussian:second:profile} over $\bar{z}$ clearly gives us $\bar{z}_{\textrm{opt}}=-32 a_1(\tilde{\phi})/a_2(\tilde{\phi})$ for $\tilde{\Delta}^{\textrm{Ga}}_{\textrm{p}}$, $\bar{z}_{\textrm{opt}}=0$ for $\tilde{\Delta}^{\textrm{Ga}}_{\textrm{m}}$, and therefore 
\begin{align}\label{optimal:photon:overlap:Gaussian:second:profile}
\tilde{\Delta}^{\textrm{Ga}}_{\textrm{p,opt}}=&\frac{\sqrt{2}\,\chi}{\sqrt{1+\chi^4}}\,\frac{e^{-4\,\frac{(\chi^2-1)^2}{\chi^4+1}\frac{\tilde{\phi}^4}{\xi(\tilde{\phi})}z_0^2}}{\sqrt{\xi(\tilde{\phi})}}e^{256\frac{a_1^2(\tilde{\phi})}{a_2(\tilde{\phi})}}\nonumber\\
\tilde{\Delta}^{\textrm{Ga}}_{\textrm{m,opt}}=&\frac{\sqrt{2}\,\chi}{\sqrt{1+\chi^4}}.
\end{align}
Finally, considering again the near-Earth regime, we have
\begin{align}\label{photon:overlap:Gaussian:approximate:second:profile}
\tilde{\Delta}^{\textrm{Ga}}_{\textrm{p,opt}}\approx&1-(1+32\tilde{\phi}^4+8\tilde{\phi}^4 z_0^2) \delta_1^2+2^9\frac{\tilde{\phi}^4 z_0^2\delta_1^4}{1+16\tilde{\phi}^4}\nonumber\\
\tilde{\Delta}^{\textrm{Ga}}_{\textrm{m,opt}}\approx&1-\delta_1^2.
\end{align}
These results lead us to conclude that, given a choice of quadratic phase, there is an error in making the na\"ive choice of redshift given by $\bar{z}=0$, since $\bar{z}_{\textrm{opt}}=-32 a_1(\tilde{\phi})/a_2(\tilde{\phi})$ here. This illustrates why we chose our specific method for identifying the genuine distortion due to gravity. 

In this specific case, however, we are working in the near-Earth regime. 
It is easy to see that $\bar{z}_{\textrm{opt}}\approx -64\frac{\tilde{\phi}^2\,z_0^2}{1+16\tilde{\phi}^4}\,\delta_1^2$ and that $|\bar{z}_{\textrm{opt}}|\leq8\,z_0^2\,\delta_1^2$. Furthermore, we note that the last term in \eqref{photon:overlap:Gaussian:approximate:second:profile} is bounded by  $512\frac{\tilde{\phi}^4 z_0^2\delta_1^4}{1+16\tilde{\phi}^4}\leq32 z_0^2\delta_1^4\ll\delta_1^2$, since $z_0^2\,\delta_1^2\ll1$ for the perturbative regime to be valid. This means that the last term that appears in $\tilde{\Delta}^{\textrm{Ga}}_{\textrm{p,opt}}$ in the result \eqref{photon:overlap:Gaussian:approximate:second:profile} can be dropped at the order that we are working here. It is also interesting to note that neither this term nor the shift $\bar{z}_{\textrm{opt}}$ can be increased (potentially unboundedly) with the phase $\tilde{\phi}$. Even more interestingly, there is a optimal phase parameter $\tilde{\phi}$ that maximizes the rigid shift (with respect to the functional dependence on $\tilde{\phi}$ itself): namely $\tilde{\phi}=1/2$. Nevertheless, in this case the optimal redshift factor $\bar{z}_{\textrm{opt}}$ is large, which means that the discrepancy between $\delta\omega_{\textrm{opt}}$ and the na\"ive value $\kappa\omega_0$ is also large.

%---------------------------------------------------------------------------------------------------%
\subsubsection{Frequency comb wavepackets}
%---------------------------------------------------------------------------------------------------%
As a second profile we choose a (modulated) frequency comb. A genuine frequency comb is a profile constituted by (infinitely) many peaks separated by a constant spacing \cite{Fortier:Baumann:2019}. In realistic situations, an envelope function will provide a cutoff to such an ideal train of peaks, and a different weight to each one of them.

\textit{Linear phase}. In this scenario, a simple normalizeable frequency comb profile reads 
\begin{align}\label{photon:wavepacket:frequency:comb}
F(z)\approx \sqrt[4]{\frac{1+\tilde{\sigma}^2}{2\,\pi}}\sum_n \frac{e^{-\frac{z^2}{4}}e^{-\frac{\tilde{\sigma}^2}{4}(z-n\,\tilde{d})^2}e^{-i\,\tilde{\phi}\,(z+z_0)}}{\sqrt{\vartheta_3\bigl(0,e^{-\frac{1}{2}\,\frac{\tilde{\sigma}^2}{1+\tilde{\sigma}^2}\tilde{d}^2}\bigr)}},
\end{align}
where we have introduced the parameter $\mu$ that dictates the width of the combs, the distance $d$ between the comb-peaks, the normalized overall width $\tilde{\sigma}:=\sigma/\mu$, the normalized distance $\tilde{d}:=d/\sigma$, and finally the Jacobi theta function $\vartheta_3(z,\tau):=\sum_{n\in\mathbb{Z}}\exp[\pi\,i\,n^2\,\tau]\exp[2\,\pi\,i\,n\,z]$.
To achieve the explicit normalization for \eqref{photon:wavepacket:frequency:comb} we have assumed that the peaks of the comb are separated enough such that their overlap is negligible. This requires setting $\tilde{d}\tilde{\sigma}\gg1$. Furthermore, we expect that $\tilde{\sigma}\gg1$ since the Gaussian envelope profile must have a (much) larger size than the individual peak. Extremely high values of $\tilde{\sigma}$ can be achieved with current technology \cite{Fang:Chen:2013}.

Lengthy algebraic calculations finally lead us to the expressions for the overlaps of interest, but the analytical expressions are not illuminating and thus we leave them in the appendices, and they can be found in \eqref{calculation:final:appendix}. Exact optimization of such expressions is not possible due to the infinite sums involved. However, we were able to find the optimal values in the near-Earth regime, which read
\begin{align}\label{calculation:final:optimal:approximate}
\tilde{\Delta}^{\textrm{Co}}_{\textrm{p,opt}}\approx&\left(1-\delta_1^2-\frac{\tilde{\sigma}^2}{2}\,\delta_1^2\right)\,e^{-\frac{2\,\delta_1^2}{\tilde{\sigma}^2}\,\tilde{\phi}^2}\nonumber\\
\tilde{\Delta}^{\textrm{Co}}_{\textrm{m,opt}}\approx&\left(1-\delta_1^2-\frac{\tilde{\sigma}^2}{2}\,\delta_1^2\right),
\end{align}
for an optimal frequency shift $\bar{z}_{\textrm{opt}}$ with value $\bar{z}_{\textrm{opt}}\approx0$.

\textit{Quadratic phase}. In this last scenario, we employ a normalizeable frequency comb profile of the form 
\begin{align}\label{photon:wavepacket:frequency:quadratic:phase:comb}
F(z)\approx \sqrt[4]{\frac{1+\tilde{\sigma}^2}{2\,\pi}}\sum_n \frac{e^{-\frac{z^2}{4}}e^{-\frac{\tilde{\sigma}^2}{4}(z-n\,\tilde{d})^2}e^{-i\,\tilde{\phi}^2\,(z+z_0)^2}}{\sqrt{\vartheta_3\bigl(0,e^{-\frac{1}{2}\,\frac{\tilde{\sigma}^2}{1+\tilde{\sigma}^2}\tilde{d}^2}\bigr)}},
\end{align}
where we assume the same conventions introduced above. Notice that the only difference with \eqref{photon:wavepacket:frequency:comb} is in the phase term, which is now quadratic.

This situation is more complex, and gives rise to a richer space of results. We start by noting that with the choice of phase above, the phase $\psi(z)\equiv-\tilde{\phi}^2(z+z_0)^2$ is a parabola centered around the origin. Since $\omega_0\gg1$, we anticipate that we will find an artificial dependence on the final results on a large $z_0$. To compensate for this, we assume that the parabola $\tilde{\phi}^2\,(z+z_0)^2$ that defines the phase is centered at or around the peak $\omega_0$ of the Gaussian envelope instead. This, effectively, means that we redefine $\tilde{\phi}^2\,(z+z_0)^2\rightarrow\tilde{\phi}^2\,(z+\delta z_0)^2$, where now $|\delta z_0|\sim\mathcal{O}(1)$ or, in other words, it is at most a small real number. Importantly, when $\delta z_0<0$ the parabola is centred toward the tail of the wavepacket, when $\delta z_0>0$ it is centred toward the front, and when $\delta z_0=0$ it is centered exactly at $\omega_0$ (or, equivalently, $z=0$). 

Optimization of the expressions over $\bar{z}$ is also cumbersome. We leave all computations to the appendix. We note that we have to introduce a constant $\zeta$ that includes the exact numerical values of the expressions and which was not obtained analytically. We estimate it to be of a few orders of magnitude, and leave its exact (or numerical) estimates to future work. Therefore, since we have that $\zeta\,\tilde{d}^2\tilde{\sigma}^2\gg1\gg\delta_1^2$, we were able to obtain
\begin{align}\label{optimal:mu:final}
\bar{z}_{\textrm{opt}}=8\,\tilde{\phi}^2\,\frac{\delta z_0-4\,\delta_1^2}{1+\Sigma}\,\delta_1,
\end{align}
where $\Sigma:=\tilde{\sigma}^2/(16\,\zeta\,\tilde{d}^2\,\tilde{\phi}^2)$.

We can study the two opposite cases of interest: (i) where $\tilde{\phi}\sim\mathcal{O}(1)$, and (ii) where $\tilde{\phi}\gg1$:
\begin{itemize}
\item[\textit{(i)}] In the first case we have $\tilde{\phi}\sim\mathcal{O}(1)$. We obtain
\begin{align}\label{calculation:final:approximate:comb:square:phase:final:case:one}
\tilde{\Delta}_{\textrm{p,opt}}^{\textrm{Co}}\approx&1-\delta_1^2-\frac{1}{2}\tilde{\sigma}^2\delta_1^2\nonumber\\
\tilde{\Delta}_{\textrm{m,opt}}^{\textrm{Co}}\approx&1-\delta_1^2-\frac{1}{2}\tilde{\sigma}^2\delta_1^2.
\end{align}
Note that the result here is independent on $\delta z_0$, and that it matches the results obtained previously.

\item[\textit{(ii)}] In the second case, i.e., where $\tilde{\phi}\gg1$, we have two possibilities. 

\textit{(ii.i)} When $\delta z_0\sim\nu\,\delta_1^2$, with $\nu$ a constant that is not large, we then have $\bar{z}_{\textrm{opt}}\approx\nu\,\delta_1^3\approx0$.  Therefore, we find the optimal overlap 
\begin{align}\label{calculation:final:approximate:comb:square:phase:final:case:two}
\tilde{\Delta}_{\textrm{p,opt}}^{\textrm{Co}}\approx&1-\delta_1^2-\frac{1}{2}\tilde{\sigma}^2\delta_1^2+16\frac{\tilde{\phi}^4}{\tilde{\sigma}^2}\delta_1^2\nonumber\\
\tilde{\Delta}_{\textrm{m,opt}}^{\textrm{Co}}\approx&1-\delta_1^2-\frac{1}{2}\tilde{\sigma}^2\delta_1^2.
\end{align}
\textit{(ii.ii)} We can also have $\delta z_0\geq1$, that is $\delta z_0$ is finite and not extremely small. Therefore $\bar{z}_{\textrm{opt}}=8\tilde{\phi}^2\,\frac{\delta z_0}{1+\Sigma}$, which finally allows us to find, with lengthy algebra, the overlap
\begin{align}\label{calculation:final:approximate:comb:square:phase:final:case:three}
\tilde{\Delta}_{\textrm{p,opt}}^{\textrm{Co}}\approx&1-\delta_1^2-\frac{1}{2}\tilde{\sigma}^2\delta_1^2+16\frac{\tilde{\phi}^4}{\tilde{\sigma}^2}\delta_1^2\nonumber\\
&-8\frac{\tilde{d}^2 \tilde{\phi}^2}{(1+\Sigma)}\left[8+\tilde{\phi}^2(1+\Sigma)+\zeta \tilde{d}^2 \tilde{\sigma}^2 (1+\Sigma)\right.\nonumber\\
&\left.+\frac{\tilde{\sigma}^4\tilde{\phi}^2}{1+\Sigma}+256 \frac{\tilde{d}^2 \tilde{\sigma}^2 \tilde{\phi}^2}{1+\Sigma}-16\zeta \tilde{d}^2 \tilde{\sigma}^2 \tilde{\phi}^4\right]\delta z_0^2 \delta_1^2\nonumber\\
\tilde{\Delta}_{\textrm{m,opt}}^{\textrm{Co}}\approx&1-\delta_1^2-\frac{1}{2}\tilde{\sigma}^2\delta_1^2.
\end{align}
It is clear that, when $\delta z_0\sim\nu\,\delta_1^2$ (i.e., we want to move back to the case (ii.i)), we can effectively set the terms proportional to $\delta z_0$ in \eqref{calculation:final:approximate:comb:square:phase:final:case:three} to zero and, as expected, the result \eqref{calculation:final:approximate:comb:square:phase:final:case:three} reduces to \eqref{calculation:final:approximate:comb:square:phase:final:case:two}.
\end{itemize}

%---------------------------------------------------------------------------------------------------%
\subsection{Comparison of the results}
%---------------------------------------------------------------------------------------------------%
We can now compare our main results. In both cases, we note that for the Gaussian and frequency-comb pure-state overlaps, there is a common term that is equal to $\tilde{\Delta}^{\textrm{Ga}}_{\textrm{m,opt}}$ and $\tilde{\Delta}^{\textrm{Co}}_{\textrm{m,opt}}$ respectively. Therefore, these two quantities can be used as benchmark for the respective pulse shapes. The deviations form this cases give us a way to compare the magnitude of the effects, i.e., the dependence of the optimal overlap on the phase. Therefore, let us introduce $\eta^{\textrm{Ga}}_{\textrm{q,opt}}:=\tilde{\Delta}^{\textrm{Ga}}_{\textrm{q,opt}}/\tilde{\Delta}^{\textrm{Ga}}_{\textrm{m,opt}}-1$ and $\eta^{\textrm{Co}}_{\textrm{q,opt}}:=\tilde{\Delta}^{\textrm{Co}}_{\textrm{q,opt}}/\tilde{\Delta}^{\textrm{Ga}}_{\textrm{m,opt}}-1$ as the relative change for both the mixed and pure state cases, for which q$=$m,p respectively. The definitions imply $\eta^{\textrm{Ga}}_{\textrm{m,opt}}=\eta^{\textrm{Co}}_{\textrm{m,opt}}=0$. This means that we look only at the relative changes for pure states compared to the mixed state scenario.

\textit{Linear phase scenario}. In the linear phase scenario we find that the relative changes read
\begin{align}\label{relative:change:linear:phase}
\eta^{\textrm{Ga}}_{\textrm{p,opt}}\approx&-2\tilde{\phi}^2\,\delta_1^2\nonumber\\
\eta^{\textrm{Co}}_{\textrm{p,opt}}\approx&-\frac{2\,\tilde{\phi}^2}{\tilde{\sigma}^2}\,\delta_1^2.
\end{align}
 \textit{Quadratic phase scenario}. In the linear phase scenario, the relative changes read
\begin{align}\label{relative:change:quadratic:phase}
\eta^{\textrm{Ga}}_{\textrm{p,opt}}\approx&-8\left\{(4+z_0^2) -64\frac{z_0^2}{1+16\tilde{\phi}^2}\delta_1^4\right\}\,\tilde{\phi}^4\,\delta_1^2\nonumber\\
\eta^{\textrm{Co}}_{\textrm{p,opt}}\approx&8\left\{2\tilde{\phi}^2-\frac{8\tilde{d}^2\tilde{\sigma}^2\delta z_0^2 }{ (1+\Sigma)}-\tilde{d}^2\tilde{\sigma}^2\delta z_0^2 \left[\tilde{\phi}^2+\zeta \tilde{d}^2 \tilde{\sigma}^2 \right.\right.\nonumber\\
+&\left.\left.\frac{\tilde{\sigma}^4\tilde{\phi}^2}{(1+\Sigma)^2}+256 \frac{\tilde{d}^2 \tilde{\sigma}^2 \tilde{\phi}^2}{(1+\Sigma)^2}-16\frac{\zeta \tilde{d}^2 \tilde{\sigma}^2 \tilde{\phi}^4}{1+\Sigma}\right]\right\}\frac{\tilde{\phi}^2\delta_1^2}{\tilde{\sigma}^2}.
\end{align}
The expressions \eqref{relative:change:linear:phase} and \eqref{relative:change:quadratic:phase} inform us on the difference between the pure state and mixed state scenarios. Therefore, they meet the goal of quantifying how a pure state with a complex relative phase in its Fock-state components can be sensitive to the curvature of spacetime. Note that one cannot take the limit $\tilde{\phi}\rightarrow\infty$, and this is true also for any of the parameters that appear in the numerator of the fractions. In fact,  recall that we have employed perturbative tools, and therefore these expressions are correct as long as \textit{all} coefficients of the perturbative parameters give rise to contributions that are smaller than one.

The expressions above also inform us on the differences between a purely Gaussian profile, and a frequency comb with the same Gaussian envelope. From our results it is clear that the relative advantage of the pure state with respect to the mixed state scenario is enhanced in the Gaussian profile when compared to the frequency comb profile. We note, however, that this behaviour will change if different profiles or phases are chosen. For these reasons, we avoid providing specific plots of the results here, which we leave to future work aimed at optimizing over initial states and phase shapes.

%---------------------------------------------------------------------------------------------------%
\section{Discussion and outlook}\label{section:discussion}
%---------------------------------------------------------------------------------------------------%
The results obtained in this work can shed light on the interplay of quantum coherence and gravity. Below, we conclude our work with some considerations regarding our results, as well as comment on potential applications and on outlook of this research direction.

%---------------------------------------------------------------------------------------------------%
\subsection{Considerations}
%---------------------------------------------------------------------------------------------------%
We have found that quantum states are deformed differently depending on their initial quantum coherence. This is not only of practical interest for potential applications to space-based protocols such as quantum communication in curved spacetime \cite{Chen:Zhang:2021}, but also for studies that focus on understanding the interplay of quantum mechanics and general relativity \cite{Rideout:Jennewein:2012,Raetzel:Wilkens:2016,Smith:Ahmadi:2019,Howl:Penrose:2019,Christodoulou:Rovelli:2019,Smith:Ahmadi:2019, Berera:2020,Smith:Ahmadi:2020}. There has been a recent surge of proposals to use genuine quantum features such as coherence and entanglement to test the nature of gravitating quantum systems \cite{Rideout:Jennewein:2012,Rosi:DAmico:2017,Carlesso:Bassi:2019,Howl:Penrose:2019}, and the present work can provide yet additional insight in this fascinating topic. In particular, we were able to quantify the difference in genuine deformation due to background curvature of the frequency spectrum that defines photonic states.
In order to univocally identify the gravitational effects on the quantum coherence present in the state, we introduced a phase parameter that dictates the unavoidable relative phase contribution to the quantum coherence of each component in the superposition that defines the state. Our results \eqref{relative:change:linear:phase} and \eqref{relative:change:quadratic:phase} explicitly show that such phase parameter governs the deviation of the pure-state case with respect to the mixed state case. When the parameter vanishes, the pure state and mixed state scenarios give the same result. This is the first strong evidence in our work that gravity affects differently mixed states and pure states with quantum coherence.

The expressions that constitute our final results for the purely Gaussian frequency profile can be obtained analytically for all redshifts. Nevertheless, we  focused all computations to the near-Earth scenario, where the redshift parameter $\chi$ has the perturbative expression $\chi=1+\delta_1+\delta_2$, and $\delta_1$ and $\delta_2$ collect the first-order and second-order corrections respectively. All results obtained in this work are proportional \textit{to lowest order} to  $\delta_1^2$. Note that this is not an approximation per se, but we find that all contributions proportional to $\delta_2$ vanish identically. This leads us to comment on this intriguing aspect. We first recall that the $\delta_1$ contributions are Newtonian in character, while the $\delta_2$ contributions depend genuinely on the curvature gradients \cite{Misner:Thorne:1973}. We know that the redshift $\chi$, by definition, depends on $\delta_1$ to lowest order. Since we have corrected for such contribution, and we have sought for the genuine distortion effects, we can conclude that we have de facto computed the \textit{genuine area overlap} of the original wavepacket with the deformed one. The area can be therefore expected to depend to lowest order on the square of a defining parameter of the system, which turns out to be $\delta_1$. This observation can be used to understand why, at lowest order, $\delta_2$ does not appear. Note that if we were to look at higher orders, the curvature corrections would be expected to appear through terms proportional, for example, to $\delta_1\delta_2$.

We end these considerations by empasizing the fact that the classical gravitational redshift for light has already been demonstrated in experiments more than half a century ago \cite{Pound:Rebka:1959}, and is indeed very well studied~\cite{Will2014}. However, we also note that the genuine wavepacket deformation has been so far lacked thorough investigation, a gap that we have therefore contributed to fill here.

%---------------------------------------------------------------------------------------------------%
\subsection{Applications and outlook}
%---------------------------------------------------------------------------------------------------%
We now proceed with discussing potential applications and outlook.

Deformations of wavepackets of light are important for applications, such as quantum communication, that rely on the quantum properties of light for their functioning. For example, in Space-based quantum repeater schemes, which have been developed in order to extend the distances at which  entanglement can be established between two users or nodes~\cite{Liorni:2021}, photons need to be stored in quantum memories after travelling through empty space~\cite{Guendogan2020}. Given our results, the efficiency of these memories will depend on the overlap $\Delta$ even after compensation of the redshift, e.g., by shifting the frequency of the photons according to equations \eqref{photon:overlap:new} and \eqref{photon:overlap:more:peakes}. More in general, quantum protocols that require entanglement sharing over long distances will be affected by issues of the type just mentioned. Any effect that occurs therefore needs to be quantified and understood in order for its mitigation \cite{Bruschi:Ralph:2014} or for its employment \cite{Bruschi:Datta:2014,Kohlrus:Bruschi:2017}. The results presented here can therefore lay the basis for measurement schemes using wavepacket genuine deformation, as well proposals for tests of gravitational effects of quantum coherence. 

Another consequence of the wavepacket deformation is that it poses an additional restriction to the rate at which photonic-based quantum communication link can operate. In fact, after a user prepares and sends a train of pulses, this will propagate and therefore will be distorted once received. Increasing distortion is expected to increase the temporal envelope of each pulse, which means that in order to achieve the same fidelity, more time needs to be introduced between two pulses compared to the hypothetical non-deformed case. This effectively reduces the time-bandwidth product due to gravitational effects. We plan to investigate these consequences in more detail in further studies.
Furthermore, this wavepacket distortion might open a new attack scheme on quantum communication protocols 
based on single photons~\cite{Bennett2020}, entangled photons~\cite{Ekert1991,Mattle1996}, mixed bases~\cite{Pavivcic2017}, and continuous variables~\cite{Weedbrook2012}. 

In order to employ the rescaled overlap defined in this work, we have applied it to simple pulse-shapes. We found that there are differences between the choices, namely a Gaussian and a frequency-comb with Gaussian envelope, and that such differences can lead to a detectable difference. However, when considering realistic implementations, it is necessary to seek the optimal input frequency profiles, focussing  in particular on the complex phase. Therefore, future work will need to address the interplay between choices of profiles that maximize the effect (i.e., minimize the overlap) and requirements due to realistic implementations. 

Modelling of realistic scenarios will additionally require the extension of our work to $3+1$ dimensions, which has been preliminary investigated already \cite{Exirifard:Culf:2021}. This can be done and analytical results can be sought, however, it is known that solving the field equations in $3+1$ curved spacetime ultimately requires numerical approaches. Since here we were interested in defining an operational way to quantify the genuine wavepacket distortion and a proof-of-principle demonstration, we leave such extensions to future studies.

%---------------------------------------------------------------------------------------------------%
\section{Conclusions}
%---------------------------------------------------------------------------------------------------%
We have studied the effects of spacetime curvature on a wavepacket of propagating light. We have introduced an operational way to distinguish between two concurrent effects induced by gravity on propagating wavepackets of light: the \textit{rigid shift} of the whole wavepacket, and the \textit{genuine wavepacket deformation}. Given that a pulse initially prepared by a sender, who is located at a different point in spacetime, will not match the one expected by the receiver, we define an overlap of the expected and received wavepackets that is optimized over all possible rigid shifts of the incoming spectrum. This guarantees that the contributions due to the rigid shift are compensated for, and that we are left with the quantification of the genuine distortion of the wavepacket due to propagation in a curved background. As wavepacket deformations will reduce the overlap between the expected and received pulses, a direct measurement of the deformation can be performed using this scheme within Hong-Ou-Mandel type experiments \cite{Hong1987, Wahl2013}, which have also been combined with changing phase profiles of the incoming photons~\cite{Specht2009}.   

We have used our scheme for detecting changes in the shape of the wavepacket to study the interplay between the initial quantum coherence present in the state of the photon and gravity. We have therefore computed the genuine distortion for two different initial states, a pure state and a completely mixed state, which have the same diagonal elements (i.e., the same probability of detecting the frequency component $\omega$ of the wavepacket). Comparison between the overlaps obtained this way shows that the initial phase in the pure state enhances the effects. In other words, it increases the distinguishability of the incoming and expected frequency profiles. We conclude that this is another evidence of the influence that gravity can play on quantum properties of physical systems.

\acknowledgements
We thank Steven J. van Enk, Christopher A. Fuchs, Jan Kohlrus, Valente Pranubon, Leila Khouri, Rosario Vunc Vedinciano, Matthias Blau and Jorma Louko for useful comments and discussions. AWS acknowledges funding by the Deutsche Forschungsgemeinschaft (DFG, German Research Foundation)
under Germany’s Excellence Strategy -- EXC-2123 QuantumFrontiers -- 390837967.
The \href{https://pixabay.com/vectors/satellite-artificial-space-data-5174090/}{satellite} of Figure~\ref{fig0} is licensed for free use by Pixabay, while the \href{https://pngimg.com/image/25352}{Earth} is licensed for non commercial use under the CC BY-NC 4.0 agreement by pngimg.com.

\bibliographystyle{apsrev4-2}
\bibliography{squishsquash.bib}

%---------------------------------------------------------------------------------------------------%
%---------------------------------------------------------------------------------------------------%
\appendix
%---------------------------------------------------------------------------------------------------%
%---------------------------------------------------------------------------------------------------%

\onecolumngrid

\newpage
%%---------------------------------------------------------------------------------------------------%
%\section{Comment}
%%---------------------------------------------------------------------------------------------------%
%
%In order to have a measurement scheme for $G_{\textrm{N}}$ we need an effect determined by a parameter
%\begin{align}
%\xi:=&\frac{M\,G_{\textrm{N}}^2\,\hbar}{c^5\,l^3}\nonumber\\
%=&\frac{M\,G_{\textrm{N}}}{c^2\,l}\frac{G_{\textrm{N}}\,\hbar}{c^3\,l^2}.
%\end{align}
%The first fraction is the redshift-like effects, while the second is a quantum and gravitational effect. This quantity would be infinitesimal. 
%
%I think, however, that we need both $G_{\textrm{N}}$ and $\hbar$ to be able to measure $G_{\textrm{N}}$.

%---------------------------------------------------------------------------------------------------%
\section{Derivation of the main formula}\label{appendix:one}
%---------------------------------------------------------------------------------------------------%
Here we compute our main result. The wavepacket is determined by the (peaked) function $F_{\omega_0}(\omega)=f_{\omega_0}(\omega)e^{-i\,\psi(\omega)}=1/\sqrt{\sigma}\tilde{f}_{\omega_0}(\omega)e^{-i\,\psi(\omega)}$, where we conveniently separate the modulus and the phase of $F_{\omega_0}(\omega)$. We start with the wavepacket overlaps 
\begin{align}\label{photon:overlap:gnac:appendix}
\Delta_{\textrm{p}}=&\left|\int_{-\infty}^{+\infty}d\omega f'_{\omega_0'}(\omega)f_{\omega_0}(\omega)\,e^{i(\psi'(\omega)-\psi(\omega))}\right|\nonumber\\
\Delta_{\textrm{m}}=&\int_{-\infty}^{+\infty}d\omega f'_{\omega_0'}(\omega)\,f_{\omega_0}(\omega).
\end{align}
These two overlaps are motivated in the main text and are defined in terms of the local quantities, such as frequencies, under control of the receiver. Here, we have $\int_{-\infty}^{+\infty}|F_{\omega_0}(\omega)|^2=1$ for normalization. We have implicitly assumed that $\sigma/\omega_0\ll1$ and that $F_{\omega_0}(\omega)=0$ effectively for all $\omega/\sigma\ll\omega_0/\sigma$. That is, we assume that the function has effectively non-zero support in the neighbourhood of the peak, and is zero at the origin, which allows us to extend the integral in the negative frequency domain. Throughout this work we need to make sure that \textit{any} transformation on the function $F_{\omega_0}(\omega)$ gives as a result another function that has the same property.
The phase $\psi(\omega)$ of the wavepacket will play a crucial role in this work.
The steps to be undertaken are described in the following.

%---------------------------------------------------------------------------------------------------%
\subsection{Deformation due to propagation in curved sapcetime}
%---------------------------------------------------------------------------------------------------%
The wavepacket that is received is deformed in a specific way as determined by general relativity. It is possible to show that 
\begin{align}\label{photon:operator:bob:appendix}
F_{\omega_0'}'(\omega)=\chi(r_{\textrm{A}},r_{\textrm{B}})\,F_{\omega_0}(\chi^2(r_{\textrm{A}},r_{\textrm{B}})\,\omega),
\end{align}
where the function $\chi(r_{\textrm{A}},r_{\textrm{B}})$ encodes the overall effect, and we have $\omega\,(h(r)-t)\rightarrow\chi^2(r_{\textrm{A}},r_{\textrm{B}})\omega\,(h(r)-\chi^{-2}(r_{\textrm{A}},r_{\textrm{B}})t)$. In the case of Bob orbiting a nonrotating spherical planet, $\chi(r_{\textrm{A}},r_{\textrm{B}})$ reads
\begin{align}\label{chi:expression:appendix}
\chi(r_{\textrm{A}},r_{\textrm{B}}):=\sqrt[4]{\frac{1-\frac{3\,M}{r_{\textrm{B}}}}{1-\frac{2\,M}{r_{\textrm{A}}}}}.
\end{align}
The gravitational redshift can be obtained for more general spacetimes following a known recipe \cite{Misner:Thorne:1973,Kohlrus:Bruschi:2017}.

%---------------------------------------------------------------------------------------------------%
\subsection{State normalization}
%---------------------------------------------------------------------------------------------------%
In this work we consider two type of states: pure states and their ``completely mixed counterpart''. Let us look at the case of a Hilbert space with a discrete energy basis $|n_{\omega_k}\rangle$ for each frequency $\omega_k$. In this case, it is possible to define, given a one-particle pure state $\hat{\rho}_{\textrm{p}}$, its completely mixed counterpart $\hat{\rho}_{\textrm{m}}$ as the diagonal state defined by the relation $\langle1_{\omega_k}|\hat{\rho}_{\textrm{m}}|1_{\omega_{k'}}\rangle\equiv\delta_{kk'}\langle1_{\omega_k}|\hat{\rho}_{\textrm{p}}|1_{\omega_{k}}\rangle$ for its elements. 
In the present situation, however, we need to generalize this relation to a continuous set of frequencies, which results problematic.

To see that this is the case let us repeat the procedure above. We start with the pure state $\hat{\rho}_{\textrm{p}}$ and na\"ively define its completely mixed counterpart $\hat{\rho}_{\textrm{m}}$ through  $\langle1_\omega|\hat{\rho}_{\textrm{m}}|1_{\omega'}\rangle=\delta(\omega-\omega')\,\langle1_\omega|\hat{\rho}_{\textrm{p}}|1_\omega'\rangle$.
It is immediate to verify that this is not a trace-class state. In fact, writing $\hat{\rho}_{\textrm{p}}=\int d\omega d\omega'\,\rho_{\textrm{p}}(\omega,\omega')|1_{\omega}\rangle\langle1_{\omega'}|$, we have that $\hat{\rho}_{\textrm{m}}=\int d\omega \rho_{\textrm{m}}(\omega)\, |1_{\omega}\rangle\langle1_{\omega}|$, and therefore
\begin{align}
\textrm{Tr}(\hat{\rho}_{\textrm{m}}):=&\int d\omega \langle1_{\omega}|\left(\int d\omega' \,\rho_{\textrm{m}}(\omega')|1_{\omega'}\rangle\langle1_{\omega'}|\right)|1_{\omega}\rangle\nonumber\\
=&\int d\omega \int d\omega'\,\rho_{\textrm{m}}(\omega')\langle1_{\omega}|1_{\omega'}\rangle\langle1_{\omega'}|1_{\omega}\rangle\nonumber\\
=&\int d\omega \int d\omega'\,\rho_{\textrm{m}}(\omega')\,\delta^2(\omega-\omega'),
\end{align}
which is ill defined since there is no proper definition for the application of squares of Dirac-delta to well-behaved functions.
 
Regardless of the problem presented here, we wish to have a mixed state counterpart to the proposed pure state
 \begin{align}\label{photon:states:first:modified:appendix}
\hat{\rho}_{\textrm{p}}(\chi)=&\chi^2\int d\omega d\omega' F_{\omega_0}(\chi^2\omega)F^*_{\omega_0}(\chi^2\omega') |1_{\omega}\rangle\langle1_{\omega'}|,
\end{align}
which is a trace-class state, with trace $\textrm{Tr}(\hat{\rho}_{\textrm{p}}(\chi))=\chi^2\int d\omega|F_{\omega_0}(\chi^2\omega)|^2=1$.

To define its mixed counterpart we note that, in practice, no realistic implementation of a detector can project exactly the state \eqref{photon:states:first:modified:appendix} on its diagonal components defined by a sharp frequency. Therefore, let us introduce the ``window projector'' $\hat{\Pi}^{(n)}:=|1^{(n)}(\theta)\rangle\langle1^{(n)}(\theta)|$, with 
 \begin{align}\label{window:state:appendix}
|1^{(n)}(\theta)\rangle:=&\int_{(n-1/2)\sigma_*}^{(n+1/2)\sigma_*}\frac{d\omega}{\sqrt{\sigma_*}}e^{i\,\theta\,\frac{\omega}{\sigma_*}}|1_{\omega}\rangle.
\end{align}
It is easy to check that $\hat{\Pi}^{(n)2}=\hat{\Pi}^{(n)}$ and that $\langle1^{(n)}(\theta)|1^{(m)}(\theta)\rangle=\delta_{nm}$.

The parameter $\sigma_*$ is crucial here. It has dimension of frequency and we assume that $\lambda:=\sigma_*/\sigma\ll1$. This means that the projector always selects very ``thin slices'' of the profile functions $F_{\omega_0}$, characterized by a width that is much smaller than the characteristic width of the peaked function $F_{\omega_0}$.

We now wish to compute the quantity $\tilde{\rho}_{nm}(\theta):=\langle1^{(n)}(\theta)|\hat{\rho}_{\textrm{p}}(\chi)|1^{(m)}(\theta)\rangle$. We have
 \begin{align}\label{window:diagonal:elements:appendix}
\tilde{\rho}_{nm}(\theta)=\chi^2\,\lambda\,\int_{n-1/2}^{n+1/2} dw \int_{m-1/2}^{m+1/2}dw'\, \tilde{f}_{\omega_0}(\chi^2\,\lambda\,w) \tilde{f}_{\omega_0}(\chi^2\,\lambda\,w')\,e^{i(\psi(\lambda\,w')-\psi(\lambda\,w))}\,e^{i\,\theta\,(w'-w)}.
\end{align}
We now ask ourselves what happens if we integrate out the angle $\theta$ from \eqref{window:diagonal:elements:appendix}. We can do this by introducing $\tilde{\rho}_{nm}:=1/\sqrt{2\pi}\int_{-\infty}^{\infty} d\theta\tilde{\rho}_{nm}(\theta)$. We find
 \begin{align}\label{window:diagonal:elements:calculation:appendix}
\tilde{\rho}_{nm}:=&\frac{1}{\sqrt{2\pi}}\int_{-\infty}^{\infty} d\theta\tilde{\rho}_{nm}(\theta)\nonumber\\
=&\frac{\chi^2\,\lambda}{\sqrt{2\pi}}\,\int_{-\infty}^{\infty} d\theta\,\int_{n-1/2}^{n+1/2} dw \int_{m-1/2}^{m+1/2}dw'\, \tilde{f}_{\omega_0}(\chi^2\,\lambda\,w) \tilde{f}_{\omega_0}(\chi^2\,\lambda\,w')\,e^{i(\psi(\lambda\,w')-\psi(\lambda\,w))}\,e^{i\,\theta\,(w'-w)}\nonumber\\
=&\chi^2\,\lambda\,\int_{n-1/2}^{n+1/2} dw \int_{m-1/2}^{m+1/2}dw'\, \tilde{f}_{\omega_0}(\chi^2\,\lambda\,w) \tilde{f}_{\omega_0}(\chi^2\,\lambda\,w')\,e^{i(\psi(\lambda\,w')-\psi(\lambda\,w))}\,\delta(w'-w),
\end{align}
which gives us
\begin{align}
\tilde{\rho}_{nm}=\left\{
\begin{array}{ll}
\chi^2\,\lambda\,\int_{n-1/2}^{n+1/2}\, \tilde{f}_{\omega_0}^2(\chi^2\,\lambda\,w) & \textrm{for}\quad n=m\\
0 & \textrm{otherwise}
\end{array}
\right.
\end{align}
Therefore, focusing only on the diagonal elements we find
\begin{align}
\tilde{\rho}_{nn}=&\chi^2\,\lambda\,\int_{n-1/2}^{n+1/2}\, \tilde{f}_{\omega_0}^2(\chi^2\,\lambda\,w)\nonumber\\
\simeq&\chi^2\,\lambda\,\int_{-1/2}^{1/2} dw \left[\tilde{f}_{\omega_0}(\chi^2\,\lambda\,n)\right]^2+\mathcal{O}(\chi^4\lambda^2)\nonumber\\
\simeq&\chi^2\,\lambda\,\tilde{f}_{\omega_0}^2(\chi^2\,\lambda\,n)+\mathcal{O}(\chi^4\lambda^2).
\end{align}
To obtain \eqref{window:diagonal:elements:calculation:appendix} we have used the perturbative expansion of the $f_{\omega_0}$ functions and the fact that $\lambda\ll1$. Therefore, we have that 
 \begin{align}\label{window:elements:final:appendix}
\tilde{\rho}_{nm}\simeq&\delta_{nm}\chi^2\,\lambda\,\tilde{f}_{\omega_0}^2(\chi^2\,\lambda\,n)+\mathcal{O}(\chi^4\lambda^2).
\end{align}
Therefore, we introduce the states $\hat{\rho}_{\textrm{m}}(\chi)$ and $\hat{\rho}_{\textrm{p}}(\chi)$ by the expressions
\begin{align}\label{photon:states:modified:appendix}
\hat{\rho}_{\textrm{m}}(\chi):=&\chi^2 \sum_n \lambda\,\tilde{f}_{\omega_0}^2(\chi^2\,\lambda\,n)\, |1^{(n)}(0)\rangle\langle1^{(n)}(0)|\nonumber\\
\hat{\rho}_{\textrm{p}}(\chi)=&\chi^2\int d\omega d\omega' F_{\omega_0}(\chi^2\omega)F^*_{\omega_0}(\chi^2\omega') |1_{\omega}\rangle\langle1_{\omega'}|,
\end{align}
which represent the mixed and pure states received by Bob.
The states that are sent can be obtained, instead, by simply setting $\chi=1$. Note that normalization of the states \eqref{photon:states:modified:appendix} is now free from the mathematical issues mentioned above. It is easy, in fact, to check that 
\begin{align}
\textrm{Tr}(\hat{\rho}_{\textrm{p}}(\chi))=&1\nonumber\\
\textrm{Tr}(\hat{\rho}_{\textrm{m}}(\chi))=&\int d\omega\chi^2 \sum_n \lambda\,\tilde{f}_{\omega_0}^2(\chi^2\,\lambda\,n)\, \langle1_\omega|1^{(n)}(0)\rangle\langle1^{(n)}(0)|1_\omega\rangle\nonumber\\
=&\chi^2 \sum_n \lambda\,\tilde{f}_{\omega_0}^2(\chi^2\,\lambda\,n)\nonumber\\
=&\int dx \tilde{f}_{\omega_0}^2(x)\nonumber\\
=&1
\end{align}
by noting that, in the continuum limit $\chi^2\,\lambda\rightarrow0$, we have $\chi^2\,\lambda\sim dx$ and $\chi^2\,\lambda\,n\sim x$.

This means that, for all purposes, the mixed state $\hat{\rho}_{\textrm{m}}(\chi)$ introduced in \eqref{photon:states:modified:appendix} achieves our goal of providing a good prescription to obtain a well-defined mixed state counterpart to the pure state $\hat{\rho}_{\textrm{p}}(\chi)$.

We can ask ourselves the following question: \textit{does the coherence of the state change due to gravity?} Intuitively, since the frequency distribution is altered, but there is no loss of information to additional degrees of freedom, we expect the mixedness of the state not to change.  To answer the question we use the \textit{purity} $\gamma$ of a quantum state $\hat{\rho}$ defined as $\gamma(\hat{\rho}):=\textrm{Tr}[\hat{\rho}^2]$, which is bounded by $0\leq\gamma(\hat{\rho})\leq1$ and is equal to unity only for pure states.
Introducing the purities $\gamma_{\textrm{m}}(\chi):=\textrm{Tr}[\hat{\rho}_{\textrm{m}}^2(\chi)]$ and $\gamma_{\textrm{p}}(\chi):=\textrm{Tr}[\hat{\rho}_{\textrm{p}}^2(\chi)]$ crucially defined \textit{in Bob's local frame}, we to see that $\gamma_{\textrm{p}}(0):=\gamma(\hat{\rho}_{\textrm{p}}(0))=1$. It is immediate to see that $\gamma_{\textrm{p}}(\chi)=\gamma_{\textrm{p}}(0)$.

In order to compute the unmodified $\gamma_{\textrm{m}}(0):=\gamma(\hat{\rho}_{\textrm{m}}(0))$ and modified $\gamma_{\textrm{m}}(\chi)$ purities of the mixed state instead, we use the modified mixed state \eqref{photon:states:modified:appendix}. We find that
\begin{align}
\gamma_{\textrm{m}}(\chi)=&\textrm{Tr}\left[\left(\chi^2 \sum_n \lambda\,\tilde{f}_{\omega_0}^2(\chi^2\,\lambda\,n)\, |1^{(n)}(0)\rangle\langle1^{(n)}(0)|\right)^2\right]\nonumber\\
=&\textrm{Tr}\left[\chi^4 \sum_{n,m} \lambda^2\,\tilde{f}_{\omega_0}^2(\chi^2\,\lambda\,n)\,\tilde{f}_{\omega_0}^2(\chi^2\,\lambda\,m)\, \langle1^{(n)}(0)|1^{(m)}(0)\rangle|1^{(n)}(0)\rangle\langle1^{(m)}(0)|\right]\nonumber\\
=&\int dx \int dx' \tilde{f}_{\omega_0}^2(x)\,\tilde{f}_{\omega_0}^2(x')\, \langle1^{(x/(\chi^2\lambda))}(0)|1^{(x'/(\chi^2\lambda))}(0)\rangle
\textrm{Tr}\left[|1^{(x/(\chi^2\lambda))}(0)\rangle\langle1^{(x'/(\chi^2\lambda))}(0)|\right]\nonumber\\
=&\int dx \tilde{f}_{\omega_0}^4(x)\, \textrm{Tr}\left[|1^{(x/(\chi^2\lambda))}(0)\rangle\langle1^{(x/(\chi^2\lambda))}(0)|\right]\nonumber\\
=&\int dx \tilde{f}_{\omega_0}^4(x)
\end{align}
and, defining $\gamma_{\textrm{m}}(0):=\int dx \tilde{f}_{\omega_0}^4(x)$, we also have that $\gamma_{\textrm{m}}(\chi)=\gamma_{\textrm{m}}(0)$. To obtain this we have used the continuum limit $\chi^2\lambda n=x$, as well as $\chi^2\lambda=dx$.
Explicitly,
\begin{align}
=&\int dx \tilde{f}_{\omega_0}^4(x)\, \textrm{Tr}\left[|1^{(x/(\chi^2\lambda))}(0)\rangle\langle1^{(x/(\chi^2\lambda))}(0)|\right]\nonumber\\
=&\int dx\,\int d\omega\int_{(n/(\chi^2\lambda)-1/2)\sigma_*}^{(n/(\chi^2\lambda)+1/2)\sigma_*}\frac{d\omega' d\omega''}{\sigma_*}\tilde{f}_{\omega_0}^4(x)\langle1_{\omega}|1_{\omega'}\rangle\langle1_{\omega''}|1_\omega\rangle\nonumber\\
=&\int dx\,\int_{(n/(\chi^2\lambda)-1/2)\sigma_*}^{(n/(\chi^2\lambda)+1/2)\sigma_*}\frac{d\omega}{\sigma_*}\tilde{f}_{\omega_0}^4(x)\nonumber\\
=&\int dx \tilde{f}_{\omega_0}^4(x).
\end{align}

%---------------------------------------------------------------------------------------------------%
\subsection{Overlap of the states}
%---------------------------------------------------------------------------------------------------%
In this work we require to compute the overlap $\Delta:=\sqrt{\mathcal{F}(\hat{\rho}',\hat{\rho})}$ between the states (wavepackets) of the two systems using the quantum fidelity $\mathcal{F}(\hat{\rho}',\hat{\rho}):=\textrm{Tr}(\sqrt{\sqrt{\hat{\rho}}\hat{\rho}'\sqrt{\hat{\rho}}})$. It is convenient to write  $F_{\omega_0}(\omega)=f_{\omega_0}(\omega)\exp[i\,\psi(\omega)]$, where $f_{\omega_0}(\omega):=|F_{\omega_0}(\omega)|$ is the modulus of the frequency distribution and $\psi(\omega)$ is its phase. Using this decomposition it is immediate to compute the overlap $\Delta_{\textrm{p}}$ for the pure state case, which reads
\begin{align}\label{photon:overlap:pure:appendix}
\Delta_{\textrm{p}}=&\chi\left|\int_{-\infty}^{+\infty}d\omega f_{\omega_0}(\chi^2 \omega)f_{\omega_0}(\omega)\,e^{i(\psi(\chi^2 \omega)-\psi(\omega))}\right|.
\end{align}
Let us quickly show how we can compute that of the mixed state case. We need to compute $\Delta_{\textrm{p}}:=\textrm{Tr}(\sqrt{\sqrt{\hat{\rho}_{\textrm{m}}(0)}\hat{\rho}_{\textrm{m}}(\chi)\sqrt{\hat{\rho}_{\textrm{m}}(0)}})$. Since we have that 
\begin{align}
\sqrt{\hat{\rho}_{\textrm{m}}(0)}=&\sqrt{\sum_n \lambda\,|F_{\omega_0}(\lambda\,n)|^2\, |1^{(n)}(0)\rangle\langle1^{(n)}(0)|}\nonumber\\
=&\sum_n \sqrt{\lambda}\,|F_{\omega_0}(\lambda\,n)||1^{(n)}(0)\rangle\langle1^{(n)}(0)|,
\end{align}
and analogously for all square-roots of diagonal states, it is not difficult to see that
\begin{align}\label{photon:overlap:mixed:appendix}
\Delta_{\textrm{m}}=&\chi\lambda\sum_n \left|F_{\omega_0}^*(\chi^2\lambda\,n)\,F_{\omega_0}(\lambda\,n)\right|\nonumber\\
=&\chi\int d\omega \left|F_{\omega_0}^*(\chi^2\omega)\,F_{\omega_0}(\omega)\right|,
\end{align}
using the continuum limit $\sigma\,\lambda\,n=\omega$ and $\sigma\,\lambda=d\omega$.

%---------------------------------------------------------------------------------------------------%
\subsection{Compensation for the ``rigid'' shift}
%---------------------------------------------------------------------------------------------------%
Physical quantum systems that are employed for concrete tasks are always localized. Therefore, the (modulus of the) functions $F_{\omega_0}(\omega)$ are assumed to be peaked and to have the functional from $f_{\omega_0}(\omega)\equiv 1/\sqrt{\sigma}\tilde{f}(\frac{\omega-\omega_0}{\sigma})$. Here $\tilde{f}$ are conveniently normalized functions as it will become clear later. We now use the relation \eqref{photon:operator:bob:appendix} and the function dependence of the wavepacket on the peak to obtain 
\begin{align}\label{photon:overlap:preliminary:appendix}
\Delta_{\textrm{p}}=&\chi\left|\int_{-\infty}^{+\infty}\frac{d\omega}{\sigma} \tilde{f}\left(\frac{\chi^2 \omega-\omega_0}{\sigma}\right)\tilde{f}\left(\frac{\omega-\omega_0}{\sigma}\right)\,e^{i(\psi(\chi^2 \omega)-\psi(\omega))}\right|\nonumber\\
\Delta_{\textrm{m}}=&\chi\int_{-\infty}^{+\infty}\frac{d\omega}{\sigma} \tilde{f}\left(\frac{\chi^2 \omega-\omega_0}{\sigma}\right)\tilde{f}\left(\frac{\omega-\omega_0}{\sigma}\right).
\end{align}
The instinct is to try to identify a translation of the frequencies as a means to compensate for the mismatch between the received packet and the expected packet. For this reason, we assume that the receiver:
\begin{itemize}
	\item[i)] can perform an \textit{arbitrary and rigid shift} of the \textit{whole} spectrum, implemented by setting $\omega\rightarrow\omega+\delta\omega$ for an arbitrary $\delta\omega$;
	\item[ii)] will apply such transformation to the \textit{incoming wavepacket only}. 
\end{itemize}
With this in mind we obtain the new overlaps
\begin{align}\label{photon:overlap:appendix}
\Delta_{\textrm{p}}=&\chi\left|\int_{-\infty}^{+\infty}\frac{d\omega}{\sigma} \tilde{f}\left(\frac{\chi^2 \omega+\chi^2\delta\omega-\omega_0}{\sigma}\right)\tilde{f}\left(\frac{\omega-\omega_0}{\sigma}\right)\,e^{i(\psi(\chi^2 \omega+\chi^2\delta\omega)-\psi(\omega))}\right|\nonumber\\
\Delta_{\textrm{m}}=&\chi\int_{-\infty}^{+\infty}\frac{d\omega}{\sigma} \tilde{f}\left(\frac{\chi^2 \omega-\omega_0}{\sigma}\right)\tilde{f}\left(\frac{\omega-\omega_0}{\sigma}\right).
\end{align}
From a formal perspective we can introduce the rescaled variable $\omega':=\omega+\omega_0$, which would centre both peaked functions around $\omega=0$ if one was not affected by propagation in a curved background. We therefore have
\begin{align}\label{photon:overlap:second:appendix}
\Delta_{\textrm{p}}=&\chi\left|\int_{-\infty}^{+\infty}\frac{d\omega'}{\sigma} \tilde{f}\left(\frac{\chi^2 \omega'+\chi^2\delta\omega+(\chi^2-1)\omega_0}{\sigma}\right)\tilde{f}\left(\frac{\omega'}{\sigma}\right)\,e^{i(\psi(\chi^2 \omega'+\chi^2\delta\omega+\chi^2\omega_0)-\psi(\omega'+\omega_0))}\right|\nonumber\\
\Delta_{\textrm{m}}=&\chi\int_{-\infty}^{+\infty}\frac{d\omega'}{\sigma} \tilde{f}\left(\frac{\chi^2 \omega'+\chi^2\delta\omega+(\chi^2-1)\omega_0}{\sigma}\right)\tilde{f}\left(\frac{\omega'}{\sigma}\right).
\end{align}
Finally, first rescaling $\omega'\rightarrow\chi\omega'$ and then introducing the new dimensionless variable $z:=\omega'/\sigma$, with the corresponding quantities $\delta z:=\delta\omega/\sigma$ and $z_0:=\omega_0/\sigma$, we have
\begin{align}\label{photon:overlap:final:appendix}
\tilde{\Delta}_{\textrm{p}}=&\left|\int_{-\infty}^{+\infty}dz \tilde{f}\left(\chi z+\bar{z}\right)\tilde{f}\left(z/\chi\right)\,e^{i(\psi(\sigma(\chi z+\bar{z}+z_0))-\psi(\sigma(z/\chi+z_0))}\right|\nonumber\\
\tilde{\Delta}_{\textrm{m}}=&\int_{-\infty}^{+\infty}dz \tilde{f}\left(\chi z+\bar{z}\right)\tilde{f}\left(z/\chi\right).
\end{align}
Here we have introduced the important quantity $\bar{z}:=\chi^2\delta z+(\chi^2-1)z_0$, which is directly related to the shift $\delta\omega$ introduced by the receiver. Note that we have changed the notation from $\Delta$ to $\tilde{\Delta}$ to highlight the change to dimensionless, shifted quantities within the integrals. Of course, the value of the integrals is independent of such change of definitions.

%---------------------------------------------------------------------------------------------------%
\subsection{Optimization over the rigid shift}
%---------------------------------------------------------------------------------------------------%
We describe our last operation. We note that the overlaps \eqref{photon:overlap:final:appendix} quantify how faithful the incoming photon is to the expected one, \textit{modulo an arbitrary rigid shift of the whole incoming wavepacket}. The final step, therefore, is to optimize the expressions \eqref{photon:overlap:final:appendix} over the shift $\bar{z}$ (which effectively means over the frequency shift $\delta\omega$). Given the value $\bar{z}_{\textrm{opt}}$ that achieves this goal, we therefore say that
\begin{itemize}
	\item[i)] The optimal dimensionless shift $\bar{z}_{\textrm{opt}}$ provides us the \textit{classical redshift} $\delta\omega_{\textrm{rs}}$ of the photon through 
	\begin{align}\label{classical:redshift:appendix}
	\delta\omega_{\textrm{rs}}:=\frac{\sigma}{\chi^2}(\bar{z}_{\textrm{opt}}-(\chi^2-1)z_0).
	\end{align}
	\item[ii)] The \textit{genuine distortion} $\tilde{\Delta}_{\textrm{m,opt}}$ and $\tilde{\Delta}_{\textrm{p,opt}}$ of the wavepacket due to gravity, defined by 
	\begin{align}\label{genuine:distortion:appendix}
	\tilde{\Delta}_{\textrm{p,opt}}:=&\left.\tilde{\Delta}_{\textrm{p}}\right|_{\bar{z}=\bar{z}_{\textrm{opt}}}\nonumber\\
	\tilde{\Delta}_{\textrm{m,opt}}:=&\left.\tilde{\Delta}_{\textrm{m}}\right|_{\bar{z}=\bar{z}_{\textrm{opt}}}.
	\end{align}
	\end{itemize}
This is our main result.

\section{Deformation effects on different wavepackets}\label{appendix:application:to:concrete:wavepackets}
%---------------------------------------------------------------------------------------------------%
Here we explicitly compute the expressions for the main quantity $\tilde{\Delta}$ in the case of: (i) a Gaussian profile; (ii) a frequency-comb with Gaussian envelope. Note that, in the following, $\tilde{F}:=\tilde{f}\exp[i\psi]$.

%---------------------------------------------------------------------------------------------------%
\subsection{Preliminary tool}
%---------------------------------------------------------------------------------------------------%
In the following we will make use of the preliminary expression
\begin{align}\label{useful:intermediate:formula:appendix}
\int_{-\infty}^{+\infty}dz\,e^{-\frac{1}{4}a_2\,z^2}\,e^{-\frac{1}{2}\,a_1\,z}\,e^{-\frac{1}{4}\,c_0}\,e^{-i\,\kappa\,z}
=&\sqrt{\frac{4\pi}{a_2}}e^{-\frac{1}{4}\,c_0}\,e^{-\frac{1}{4}\frac{a_0\,a_2-a_1^2}{a_2}}\,e^{-\frac{\kappa^2}{a_2}}\,e^{i\,\kappa\,\frac{a_1}{a_2}}.
\end{align}
This will allow us to compute the $\tilde{\Delta}$ quantities for all of our cases. In particular, it will allow us to do so for the case where the phase $\psi(z)$ is present, i.e., the pure-state case, which can be immediately adapted to the mixed-state case by setting $\psi=0$. 

%---------------------------------------------------------------------------------------------------%
\subsection{Gaussian wavepackets}
%---------------------------------------------------------------------------------------------------%
We start by considering Gaussian wavepackets of light. It has been argued that Gaussian distributions are resilient to modifications of the type discussed in this work \cite{Kish:Ralph:2016}. For this reason, we can already anticipate that the choice of such functions will result into small deformations. We will consider wavepackets with two different frequency-dependent phases. 

%---------------------------------------------------------------------------------------------------%
\subsubsection{Gaussian wavepackets: linear phase}
%---------------------------------------------------------------------------------------------------%
The first wavepacket is defined through the normalized Gaussian with a phase that is linear in the frequency. In our normalized variable $z$ this reads $\tilde{F}(z)=1/\sqrt[4]{2\,\pi}\exp[-z^2/4-i\,\tilde{\phi}\,(z+z_0)]$, which allows us to easily compute the overlaps $\tilde{\Delta}^{\textrm{Ga}}$ as
\begin{align}\label{photon:overlap:Gaussian:appendix}
\tilde{\Delta}^{\textrm{Ga}}_{\textrm{p}}=&\frac{\sqrt{2}\,\chi}{\sqrt{1+\chi^4}}\,e^{-\frac{\bar{z}}{\chi^4+1}}\,e^{-\frac{(\chi^2-1)^2}{\chi^4+1}\tilde{\phi}^2}\nonumber\\
\tilde{\Delta}^{\textrm{Ga}}_{\textrm{m}}=&\frac{\sqrt{2}\,\chi}{\sqrt{1+\chi^4}}\,e^{-\frac{\bar{z}}{\chi^4+1}}.
\end{align}
Note that the results above are independent of $z_0$. This occurs because it induces a global phase contribution for the mode of the photon. Instead, the overalps depend on the arbitrary rescaled shift $\bar{z}$.

Optimization over $\bar{z}$ clearly gives $\bar{z}_{\textrm{opt}}=0$, which means that the optimal $\tilde{\Delta}^{\textrm{Ga}}$ quantities read
\begin{align}\label{optimal:photon:overlap:Gaussian:appendix}
\tilde{\Delta}^{\textrm{Ga}}_{\textrm{p,opt}}=&\frac{\sqrt{2}\,\chi}{\sqrt{1+\chi^4}}\,e^{-\frac{(\chi^2-1)^2}{\chi^4+1}\tilde{\phi}^2}\nonumber\\
\tilde{\Delta}^{\textrm{Ga}}_{\textrm{m,opt}}=&\frac{\sqrt{2}\,\chi}{\sqrt{1+\chi^4}}.
\end{align}
When we consider the near-Earth regime, we then have
\begin{align}\label{photon:overlap:Gaussian:approximate:appendix}
\tilde{\Delta}^{\textrm{Ga}}_{\textrm{p}}\approx&1-(1+2\tilde{\phi}^2)\,\delta_1^2\nonumber\\
\tilde{\Delta}^{\textrm{Ga}}_{\textrm{m}}\approx&1-\delta_1^2.
\end{align}

%---------------------------------------------------------------------------------------------------%
\subsubsection{Gaussian wavepackets: quadratic phase}
%---------------------------------------------------------------------------------------------------%
We can choose a different phase profile for our Gaussian and therefore employ the Gaussian $\tilde{F}(z)=1/\sqrt[4]{2\,\pi}\exp[-z^2/4-i\,\tilde{\phi}\,(z+z_0)^2]$ in normalized and shifted coordinates. Then, we find
\begin{align}\label{photon:overlap:Gaussian:second:profile:appendix}
\tilde{\Delta}^{\textrm{Ga}}_{\textrm{p}}=&\frac{\sqrt{2}\,\chi}{\sqrt{1+\chi^4}}\,\frac{e^{-4\,\frac{(\chi^2-1)^2}{\chi^4+1}\frac{\tilde{\phi}^2}{\xi(\tilde{\phi})}z_0^2}}{\sqrt{\xi(\tilde{\phi})}}e^{-16\,a_1(\tilde{\phi}) \bar{z}-\frac{1}{4}\,a_2(\tilde{\phi}) \bar{z}^2}\nonumber\\
\tilde{\Delta}^{\textrm{Ga}}_{\textrm{m}}=&\frac{\sqrt{2}\,\chi}{\sqrt{1+\chi^4}}e^{-\frac{1}{4}\,a_2(0) \bar{z}^2},
\end{align}
where we have introduced 
\begin{align}
\xi(\tilde{\phi}):=&1+16\tilde{\phi}^2\frac{(\chi^4-1)^2}{(\chi^4+1)^2}\nonumber\\
a_1(\tilde{\phi}):=&\chi^2\frac{(\chi^2-1)^2}{(\chi^4+1)^2}\frac{\tilde{\phi}}{\xi}z_0\nonumber\\
a_2(\tilde{\phi}):=&\frac{1+16\tilde{\phi}^2}{\chi^4+1}\frac{1}{\xi}
\end{align} 
as $\tilde{\phi}$-dependent functions.
Optimizing over $\bar{z}$ gives us
\begin{align}\label{optimal:photon:overlap:Gaussian:second:profile:appendix}
\tilde{\Delta}^{\textrm{Ga}}_{\textrm{p,opt}}=&\frac{\sqrt{2}\,\chi}{\sqrt{1+\chi^4}}\,\frac{e^{-4\,\frac{(\chi^2-1)^2}{\chi^4+1}\frac{\tilde{\phi}^2}{\xi(\tilde{\phi})}z_0^2}}{\sqrt{\xi(\tilde{\phi})}}e^{256\frac{a_1^2(\tilde{\phi})}{a_2(\tilde{\phi})}}\nonumber\\
\tilde{\Delta}^{\textrm{Ga}}_{\textrm{m,opt}}=&\frac{\sqrt{2}\,\chi}{\sqrt{1+\chi^4}},
\end{align}
with the optimal value $\bar{z}_{\textrm{opt}}=-32\frac{a_1^2(\tilde{\phi})}{a_2(\tilde{\phi})}$.

Finally, considering again the near-Earth regime, we have
\begin{align}\label{photon:overlap:Gaussian:approximate:second:profile:appendix}
\tilde{\Delta}^{\textrm{Ga}}_{\textrm{p,opt}}\approx&1-(1+32\tilde{\phi}^2+8\tilde{\phi}^2z_0^2)\,\delta_1^2+\frac{512\,\tilde{\phi}^2}{1+16\tilde{\phi}^2}\,z_0^2\,\delta_1^4\nonumber\\
\tilde{\Delta}^{\textrm{Ga}}_{\textrm{m,opt}}\approx&1-\delta_1^2,
\end{align}
to be supplemented with the expression $\bar{z}_{\textrm{opt}}=-\frac{64\,\tilde{\phi}^2}{1+16\tilde{\phi}^2}\,z_0^2\,\delta_1^2$. 

%---------------------------------------------------------------------------------------------------%
\subsection{Frequency comb wavepackets}\label{appendix:frequency:comb}
%---------------------------------------------------------------------------------------------------%
Here we consider a frequency comb, which is a profile constituted by many peaks separated by a constant spacing \cite{Fortier:Baumann:2019}. 
The normalizeable frequency comb profile as a function of the shifted dimensionless frequency $z$ reads 
\begin{align}\label{photon:wavepacket:frequency:comb:appendix}
\tilde{F}(z)\approx \sqrt[4]{\frac{1+\tilde{\sigma}^2}{8\,\pi}}\sum_n \frac{e^{-\frac{z^2}{4}}e^{-\frac{\tilde{\sigma}^2}{4}(z-n\,\tilde{d})^2}e^{-i\,\tilde{\phi}\,(z+z_0)}}{\sqrt{\vartheta_3\left(0,e^{-\frac{1}{2}\,\frac{\tilde{\sigma}^2}{1+\tilde{\sigma}^2}\tilde{d}^2}\right)}},
\end{align}
where $d$ determines the distance between the comb-peaks, $\mu$ is the width of the combs, and $\vartheta_3(z,\tau)$ is the Jacobi theta function. We have also defined $\tilde{\sigma}:=\sigma/\mu$ and $\tilde{d}:=d/\sigma$ for convenience of presentation. 

To achieve this explicit normalization we have assumed that the combs are separated enough such that their overlap is negligible. This requires setting $\tilde{d}\tilde{\sigma}\gg1$. Furthermore, we expect that $\tilde{\sigma}\gg1$ since the Gaussian envelope profile has a (much) larger size than the individual comb. Extremely high values of $\tilde{\sigma}$ can be achieved with current technology \cite{Fang:Chen:2013}. 
Therefore, we started from the general function
\begin{align}\label{photon:wavepacket:frequency:comb:start:appendix}
\tilde{F}(z)= C\sum_n e^{-\frac{z^2}{4}}e^{-\frac{\tilde{\sigma}^2}{4}(z+z_0-n\,\tilde{d})^2}e^{-i\,\psi(z+z_0)},
\end{align}
to compute the constant $C$ such that $\int_{-\infty}^{+\infty}|\tilde{F}(z)|^2=1$. Here $\psi(z)$ is an arbitrary $z$-dependent phase. We have
\begin{align}\label{photon:wavepacket:frequency:comb:normalization:appendix}
1=\int_{-\infty}^{+\infty}dz|\tilde{F}(z)|^2=&C^2\sum_{n,m\in\mathbb{Z}}e^{-\frac{1}{8}\frac{\tilde{\sigma}^2}{1+\tilde{\sigma}^2}\tilde{d}^2\left[2(1+\tilde{\sigma}^2)(n^2+m^2)-\tilde{\sigma}^2(n+m)^2\right]}\int_{-\infty}^{+\infty}dz\,e^{-\frac{1+\tilde{\sigma}^2}{2}\left(z-\frac{1}{2}(n+m)\frac{\tilde{\sigma}^2}{1+\tilde{\sigma}^2}\tilde{d}^2\right)^2}\nonumber\\
=&C^2\sqrt{\frac{2\,\pi}{1+\tilde{\sigma}^2}}\sum_{n,m\in\mathbb{Z}}e^{-\frac{1}{8}\frac{\tilde{\sigma}^2}{1+\tilde{\sigma}^2}\tilde{d}^2\left[2(n^2+m^2)+\tilde{\sigma}^2(n-m)^2\right]}\nonumber\\
=&C^2\sqrt{\frac{2\,\pi}{1+\tilde{\sigma}^2}}\left[1+2\sum_{n,m=1}^{+\infty}e^{-\frac{1}{4}(n^2+m^2)\frac{\tilde{\sigma}^2}{1+\tilde{\sigma}^2}\tilde{d}^2}\left[e^{-\frac{1}{8}(n+m)^2\frac{\tilde{\sigma}^4}{1+\tilde{\sigma}^2}\tilde{d}^2}+e^{-\frac{1}{8}(n-m)^2\frac{\tilde{\sigma}^4}{1+\tilde{\sigma}^2}\tilde{d}^2}\right]\right]\nonumber\\
=&2\,C^2\sqrt{\frac{2\,\pi}{1+\tilde{\sigma}^2}}\left[1+2\sum_{n\in\mathbb{N}}e^{-\frac{1}{2}n^2\frac{\tilde{\sigma}^2}{1+\tilde{\sigma}^2}\tilde{d}^2}+4\,\sum_{n=0,k=1}\ldots\right]\nonumber\\
\approx& 2\,C^2\sqrt{\frac{2\,\pi}{1+\tilde{\sigma}^2}}\vartheta_3\left(0,e^{-\frac{1}{2}\,\frac{\tilde{\sigma}^2}{1+\tilde{\sigma}^2}\tilde{d}^2}\right),
\end{align}
where all other contributions can be neglected (and included if necessary). To obtain the above we assumed that there is a particular value $n_*$ such that $n_*\,\tilde{d}=z_0$, that is, that the $n_*$th peak is aligned with the peak $z_0$. This is not crucial but convenient, and this condition can be relaxed if necessary. Furthermore, this condition allowed us to shift the sum such that $n=0\rightarrow n=n_*$. In turn, this allows us to find the normalization constant $C$ as 
\begin{align}
C=\sqrt{\frac{1+\tilde{\sigma}^2}{8\,\pi}}\frac{1}{\sqrt{\vartheta_3\left(0,e^{-\frac{1}{2}\,\frac{\tilde{\sigma}^2}{1+\tilde{\sigma}^2}\tilde{d}^2}\right)}},
\end{align}
which justifies the choice of function \eqref{photon:wavepacket:frequency:comb:appendix}.

%---------------------------------------------------------------------------------------------------%
\subsubsection{Frequency comb wavepackets: linear phase}
%---------------------------------------------------------------------------------------------------%
As done for the pure Gaussian profile, we here start with the the linear phase case $\psi(z)=\tilde{\phi} (z+z_0)$. We have
\begin{align}\label{calculation:start:appendix}
\tilde{\Delta}_{\textrm{p}}=&\left|\int_{-\infty}^{+\infty}dz\,\tilde{f}^*(\chi\,z+\bar{z})\tilde{f}(z/\chi)\,e^{i\,\tilde{\phi}\,(\chi\,z+\bar{z}+z_0)}\,e^{-i\,\tilde{\phi}\,(z/\chi+z_0)}\right|\nonumber\\
=&\sqrt{\frac{1+\tilde{\sigma}^2}{8\,\pi}}\frac{1}{\vartheta_3\bigl(0,\exp\bigl[-\frac{1}{2}\,\frac{\tilde{\sigma}^2}{1+\tilde{\sigma}^2}\tilde{d}^2\bigr]\bigr)}\left|\sum_{n,m\in\mathbb{Z}}\int_{-\infty}^{+\infty}dz\,e^{-\frac{(\chi\,z+\bar{z})^2}{4}}\,e^{-\frac{(z/\chi)^2}{4}}\,e^{-\frac{\tilde{\sigma}^2}{4}(\chi\,z+\bar{z}+z_0-n\,\tilde{d})^2}\,e^{-\frac{\tilde{\sigma}^2}{4}(z/\chi+z_0-m\,\tilde{d})^2}\right.\nonumber\\
&\left.\times e^{i\,\tilde{\phi}\,(\chi\,z+\bar{z}+z_0)}\,e^{-i\,\tilde{\phi}\,(z/\chi+z_0)}\right|\nonumber\\
=&\sqrt{\frac{2\,\chi^2}{1+\chi^4}}\frac{1}{\vartheta_3\bigl(0,\exp\bigl[-\frac{1}{2}\,\frac{\tilde{\sigma}^2}{1+\tilde{\sigma}^2}\tilde{d}^2\bigr]\bigr)}\left|\sum_{n,m\in\mathbb{Z}}e^{-\frac{1}{4}\frac{a_0\,a_2-a_1^2}{a_2}}\,e^{-\frac{\kappa^2}{a_2}}\,e^{i\,\kappa\,\frac{a_1}{a_2}}\right|,
\end{align}
where we need to introduce the ($n$- and $m$-dependent) coefficients for this case
\begin{align}\label{coefficients:linear:phase:appendix}
a_2:=&(1+\tilde{\sigma}^2)\frac{\chi^4+1}{\chi^2}\nonumber\\
a_1:=&\chi\,\bar{z}+\chi\,\tilde{\sigma}^2\,(\bar{z}-n\,\tilde{d})-\frac{1}{\chi}\tilde{\sigma}^2\,m\,\tilde{d}\nonumber\\
a_0:=&\bar{z}^2+\tilde{\sigma}^2\,(\bar{z}-n\,\tilde{d})^2+\tilde{\sigma}^2\,m^2\,\tilde{d}^2\nonumber\\
\kappa:=&\frac{\chi^2-1}{\chi}\,\tilde{\phi}.
\end{align}
Note that we have again used the fact that we assume that there exists an $n_*$ such that $n_*\tilde{d}=z_0$, and we shift the summation -- therefore effectively setting $z_0=0$.

Rearranging and manipulating terms, lengthy algebra gives
\begin{align}\label{calculation:middle:middle:appendix}
\tilde{\Delta}_{\textrm{p}}=&\Gamma\,\left|\sum_{n,m\in\mathbb{Z}}\left[e^{-\frac{1}{4}\tilde{\sigma}^2(n^2+m^2)\,\tilde{d}^2}\,
e^{\frac{1}{4}\frac{\chi^2}{\chi^4+1}\frac{\tilde{\sigma}^4}{1+\tilde{\sigma}^2}(\chi\,n+m/\chi)^2\,\tilde{d}^2}\,
e^{\frac{1}{2}\tilde{\sigma}^2\,n\,\tilde{d}\,\bar{z}}\,
e^{-\frac{1}{2}\frac{\chi^3}{\chi^4+1}\tilde{\sigma}^2(\chi\,n+m/\chi)\,\tilde{d}\bar{z}}\,e^{-i\,\frac{\tilde{\sigma}^2}{1+\tilde{\sigma}^2}\,\frac{\chi(\chi^2-1)}{\chi^4+1}(\chi\,n+m/\chi)\,\tilde{d}\,\tilde{\phi}}\right]\right|,
\end{align}
where the proportionality constant is
\begin{align}\label{proportionality:constant:appendix}
\Gamma:=\sqrt{\frac{2\,\chi^2}{1+\chi^4}}\frac{e^{-\frac{1+\tilde{\sigma}^2}{4}\frac{\bar{z}^2}{\chi^4+1}}\,
e^{-\frac{(\chi^2-1)^2}{\chi^4+1}\frac{\tilde{\phi}^2}{1+\tilde{\sigma}^2}}}
{\vartheta_3\bigl(0,\exp\bigl[-\frac{1}{2}\,\frac{\tilde{\sigma}^2}{1+\tilde{\sigma}^2}\tilde{d}^2\bigr]\bigr)}.
\end{align}
As already done when working to obtain the normalization constant, we see that the major contributions occur for $n=m$. Therefore, we can now approximate \eqref{calculation:middle:middle:appendix} as
\begin{align}\label{calculation:final:appendix}
\tilde{\Delta}_{\textrm{p}}\approx&\Gamma\,\left|\sum_{n\in\mathbb{Z}}
e^{-\frac{1}{2}\frac{\tilde{\sigma}^2}{1+\tilde{\sigma}^2}\left(1+\frac{\tilde{\sigma}^2}{2}\frac{(\chi^2-1)^2}{\chi^4+1}\right)\,\tilde{d}^2\,n^2}\,
e^{-\frac{1}{2}\frac{\tilde{\sigma}^2}{1+\tilde{\sigma}^2}\,\frac{\chi^2-1}{\chi^4+1}\left(\bar{z}+2\,i\,\frac{\chi^2+1}{\chi^2}\,\tilde{\phi}\right)\,\tilde{d}\,n}\right|.
\end{align}
Note that it is immediate to check that, for $\chi=1$, $\bar{z}=z_0=\tilde{\phi}=0$ we recover $\tilde{\Delta}_{\textrm{p}}^{\textrm{Ga}}$ as expected.

We now want to optimize the expression \eqref{calculation:final:appendix} with respect to $\bar{z}$. It is very difficult to do so analytically using the general expression \eqref{calculation:final:appendix}. Therefore, we resort to use the approximation $\chi\approx1+\delta_1$, and obtain
\begin{align}\label{calculation:final:approximate:appendix}
\tilde{\Delta}_{\textrm{p}}\approx&(1-\delta_1^2)\frac{e^{-\frac{1+\tilde{\sigma}^2}{8}(1-2\delta_1)\bar{z}^2}\,e^{-\frac{2\,\delta_1^2}{1+\tilde{\sigma}^2}\,\tilde{\phi}^2}}{\vartheta_3\bigl(0,\exp\bigl[-\frac{1}{2}\,\frac{\tilde{\sigma}^2}{1+\tilde{\sigma}^2}\tilde{d}^2\bigr]\bigr)}\,\left|\sum_{n\in\mathbb{Z}}
e^{-\frac{1}{2}\frac{\tilde{\sigma}^2}{1+\tilde{\sigma}^2}\left(1+\tilde{\sigma}^2\,\delta_1^2\right)\tilde{d}^2\,n^2}\,
e^{-\frac{1}{2}\frac{\tilde{\sigma}^2}{1+\tilde{\sigma}^2}\,\delta_1^2\,\left(\bar{z}+4\,i\,\tilde{\phi}\right)\,\tilde{d}\,n}\right|.
\end{align}
We now further apply the property $\tilde{\sigma}\gg1$ to obtain
\begin{align}\label{calculation:final:approximate:two:appendix}
\tilde{\Delta}_{\textrm{p}}\approx&(1-\delta_1^2)\frac{e^{-\frac{\tilde{\sigma}^2}{8}(1+1/\tilde{\sigma}^2-2\delta_1)\bar{z}^2}\,
e^{-\frac{2\,\delta_1^2}{\tilde{\sigma}^2}\,\tilde{\phi}^2}}{\vartheta_3\bigl(0,\exp\bigl[-\frac{1}{2}\,(1+1/\tilde{\sigma}^2)\tilde{d}^2\bigr]\bigr)}\,\left|\sum_{n\in\mathbb{Z}}
e^{-\frac{1}{2}\left(1+1/\tilde{\sigma}^2+\tilde{\sigma}^2\,\delta_1^2\right)\tilde{d}^2\,n^2}\,
e^{-\frac{1}{2}\delta_1^2\,\left(\bar{z}+4\,i\,\tilde{\phi}\right)\,\tilde{d}\,n}\right|.
\end{align}
Note that $\delta_1^2\ll1$ and, therefore, the last term in \eqref{calculation:final:approximate:appendix} will always be negligible. In fact, within the modulus operation this term will give a correction of order $\delta_1^4$ which we can completely ignore. Even if $\tilde{\phi}$ and $\bar{z}$ are large, we have that $\delta_1^2\,\tilde{\phi}\ll1$ and $\delta_1^2\,\bar{z}\ll1$ for the perturbative regime to be valid. This implies that the terms $\exp[-\frac{\tilde{\sigma}^2}{8}\bar{z}^2]$ and $\exp[-\frac{1}{2}\tilde{d}^2\,n^2]$ will be exponentially small, effectively cancelling any contribution from such last term. Therefore, we have
\begin{align}\label{calculation:final:final:approximate:appendix}
\tilde{\Delta}_{\textrm{p}}\approx&(1-\delta_1^2)\frac{\vartheta_3\bigl(0,\exp\bigl[-\frac{1}{2}\left(1+1/\tilde{\sigma}^2+\tilde{\sigma}^2\,\delta_1^2\right)\tilde{d}^2\bigr]\bigr)}{\vartheta_3\bigl(0,\exp\bigl[-\frac{1}{2}\,(1+1/\tilde{\sigma}^2)\tilde{d}^2\bigr]\bigr)}
e^{-\frac{\tilde{\sigma}^2}{8}(1+1/\tilde{\sigma}^2-2\delta_1)\bar{z}^2}\,e^{-\frac{2\,\delta_1^2}{\tilde{\sigma}^2}\,\tilde{\phi}^2}
\end{align}
We now wish to optimize the expression \eqref{calculation:final:final:approximate:appendix} with respect to $\bar{z}$, which trivially gives $\bar{z}_{\textrm{opt.}}=0$. 
Finally, using the expansion $\vartheta(0,\exp[-x])$ for $x\ll1$ of the Jacobi theta function \cite{Berndt:Byungchan:2011} since $\tilde{d}\ll1$, we can write
\begin{align}\label{calculation:final:optimal:approximate:appendix}
\tilde{\Delta}^{\textrm{Co}}_{\textrm{m,opt}}\approx&\left(1-\frac{\tilde{\sigma}^2}{2}\,\delta_1^2\right)\nonumber\\
\tilde{\Delta}^{\textrm{Co}}_{\textrm{p,opt}}\approx&\left(1-\frac{\tilde{\sigma}^2}{2}\,\delta_1^2\right)\,e^{-\frac{2\,\delta_1^2}{\tilde{\sigma}^2}\,\tilde{\phi}^2}.
\end{align}

%---------------------------------------------------------------------------------------------------%
\subsubsection{Frequency comb wavepackets: quadratic phase}
%---------------------------------------------------------------------------------------------------%
We proceed to consider a quadratic phase for our frequency comb scenario. In this case we have $\psi(z)=\tilde{\phi}^2 (z+z_0)^2$, and therefore
\begin{align}\label{calculation:start:appendix}
\tilde{\Delta}_{\textrm{p}}=&\left|\int_{-\infty}^{+\infty}dz\,\tilde{f}^*(\chi\,z+\bar{z})\tilde{f}(z/\chi)\,e^{i\,\tilde{\phi}^2\,(\chi\,z+\bar{z}+z_0)^2}\,e^{-i\,\tilde{\phi}^2\,(z/\chi+z_0)^2}\right|\nonumber\\
=&\sqrt{\frac{1+\tilde{\sigma}^2}{8\,\pi}}\frac{1}{\vartheta_3\bigl(0,\exp\bigl[-\frac{1}{2}\,\frac{\tilde{\sigma}^2}{1+\tilde{\sigma}^2}\tilde{d}^2\bigr]\bigr)}\left|\sum_{n,m\in\mathbb{Z}}\int_{-\infty}^{+\infty}dz\,e^{-\frac{(\chi\,z+\bar{z})^2}{4}}\,e^{-\frac{(z/\chi)^2}{4}}\,e^{-\frac{\tilde{\sigma}^2}{4}(\chi\,z+\bar{z}+z_0-n\,\tilde{d})^2}\,e^{-\frac{\tilde{\sigma}^2}{4}(z/\chi+z_0-m\,\tilde{d})^2}\right.\nonumber\\
&\left.\times e^{\frac{\chi^4-1}{\chi^2}\,i\,\tilde{\phi}^2\,z^2}\,e^{2\,(\chi\,\bar{z}+\frac{\chi^2-1}{\chi}\,z_0)\,i\,\tilde{\phi}^2\,z}\right|\nonumber\\
=&\sqrt{\frac{2\,\chi^2}{1+\chi^4}}\frac{1}{\vartheta_3\bigl(0,\exp\bigl[-\frac{1}{2}\,\frac{\tilde{\sigma}^2}{1+\tilde{\sigma}^2}\tilde{d}^2\bigr]\bigr)}\left|\sum_{n,m\in\mathbb{Z}}e^{-\frac{1}{4}\frac{a_0\,a_2-a_1^2}{a_2}}\,e^{-\frac{\kappa^2}{a_2}}\,e^{i\,\kappa\,\frac{a_1}{a_2}}\right|.
\end{align}
Here, we introduce the ($n$- and $m$-dependent) coefficients that read
\begin{align}\label{coefficients:linear:phase:appendix}
a_2:=&(1+\tilde{\sigma}^2)\frac{\chi^4+1}{\chi^2}\xi(\tilde{\phi})\nonumber\\
a_1:=&\chi\,\bar{z}+\chi\,\tilde{\sigma}^2\,(\bar{z}-n\,\tilde{d})-\frac{1}{\chi}\tilde{\sigma}^2\,m\,\tilde{d}\nonumber\\
a_0:=&\bar{z}^2+\tilde{\sigma}^2\,(\bar{z}-n\,\tilde{d})^2+\tilde{\sigma}^2\,m^2\,\tilde{d}^2\nonumber\\
\kappa:=&2\,\left(\chi\,\bar{z}+\frac{\chi^2-1}{\chi}\,z_0\right)\,\tilde{\phi}^2\nonumber\\
\xi(\tilde{\phi}):=&1-4\,\frac{\chi^4-1}{\chi^4+1}\,\frac{\tilde{\phi}^2}{1+\tilde{\sigma}^2}\,i\nonumber\\
|\xi(\tilde{\phi})|^2=&1+16\,\frac{(\chi^4-1)^2}{(\chi^4+1)^2}\,\frac{\tilde{\phi}^4}{(1+\tilde{\sigma}^2)^2}.
\end{align}
Note that we have again used the fact that we assume that there exists an $n_*$ such that $n_*\tilde{d}=z_0$ and we shift the summation -- therefore effectively setting $z_0=0$.

Rearranging as done before, lengthy algebra allows us to obtain
\begin{align}\label{calculation:middle:appendix}
\tilde{\Delta}_{\textrm{p}}=&\Gamma\,\left|\sum_{n,m\in\mathbb{Z}}\left[e^{-\frac{1}{4}\tilde{\sigma}^2(n^2+m^2)\,\tilde{d}^2}\,
e^{\frac{1}{4}\frac{\chi^2}{\chi^4+1}\frac{\tilde{\sigma}^4}{1+\tilde{\sigma}^2}(\chi\,n+m/\chi)^2\,\frac{\xi(\tilde{\phi})^*}{|\xi(\tilde{\phi})|^2}\tilde{d}^2}\,
e^{\frac{1}{2}\tilde{\sigma}^2\,n\,\tilde{d}\,\bar{z}}\,
e^{-\frac{1}{2}\frac{\chi^3}{\chi^4+1}\tilde{\sigma}^2(\chi\,n+m/\chi)\,\frac{\xi(\tilde{\phi})^*}{|\xi(\tilde{\phi})|^2}\tilde{d}\bar{z}}\right.\right.\nonumber\\
&\left.\left.e^{-2\,i\,\frac{\tilde{\sigma}^2}{1+\tilde{\sigma}^2}\,\frac{\chi^2}{\chi^4+1}\left(\chi\bar{z}+\frac{\chi^2-1}{\chi}z_0\right)(\chi\,n+m/\chi)\,\frac{\xi(\tilde{\phi})^*}{|\xi(\tilde{\phi})|^2}\,\tilde{d}\,\tilde{\phi}^2}\right]\right|,
\end{align}
where the proportionality constant is
\begin{align}\label{proportionality:constant:appendix}
\Gamma:=&\sqrt{\frac{2\,\chi^2}{1+\chi^4}}\frac{
e^{-\frac{1}{4}\frac{1+\tilde{\sigma}^2}{\chi^4+1}\left(1+16\frac{\tilde{\phi}^4}{(1+\tilde{\sigma}^2)^2}\right)\,\frac{\bar{z}^2}{|\xi(\tilde{\phi})|^2}}\,
e^{-16\frac{\chi^2(\chi^2-1)}{(\chi^4+1)^2}\frac{\tilde{\phi}^4}{(1+\tilde{\sigma}^2)|\xi(\tilde{\phi})|^2}z_0\bar{z}}
e^{-4\frac{(\chi^2-1)^2}{\chi^4+1}\frac{\tilde{\phi}^4}{(1+\tilde{\sigma}^2)}\frac{z_0^2}{|\xi(\tilde{\phi})|^2}}
}
{\vartheta_3\bigl(0,\exp\bigl[-\frac{1}{2}\,\frac{\tilde{\sigma}^2}{1+\tilde{\sigma}^2}\tilde{d}^2\bigr]\bigr)}
\end{align}
It is easy to check that, for $\chi=1$, $\bar{z}=z_0=\tilde{\phi}=0$ we recover $\tilde{\Delta}_{\textrm{p}}^{\textrm{Ga}}$ as expected.

Once more, as done in the case of obtaining the normalization constant, we see that the major contributions occur for $n=m$. Therefore, we can now approximate \eqref{calculation:middle:appendix} as
\begin{align}\label{calculation:final:comb:square:phase:appendix}
\tilde{\Delta}_{\textrm{p}}\approx&\Gamma\,\left|\sum_{n\in\mathbb{Z}}
\left[e^{-\frac{1}{2}\frac{\tilde{\sigma}^2}{1+\tilde{\sigma}^2}\left(1+\frac{1}{2}\tilde{\sigma}^2\frac{(\chi^2-1)^2}{\chi^4+1}-2\frac{\tilde{\sigma}^2}{(1+\tilde{\sigma}^2)^2}\frac{(\chi^2+1)^2(\chi^4-1)^2}{(\chi^4+1)^2}\frac{\tilde{\phi}^4}{|\xi(\tilde{\phi})|^2}-2\,i\,\frac{\tilde{\sigma}^2}{1+\tilde{\sigma}^2}\frac{(\chi^2+1)^2(\chi^4-1)^2}{(\chi^4+1)^2}\frac{\tilde{\phi}^2}{|\xi(\tilde{\phi})|^2}\right)\,n^2\,\tilde{d}^2}\right.\right.\nonumber\\
&e^{-\tilde{\sigma}^2\frac{(\chi^2-1)}{\chi^4+1}\left(1-16\frac{1}{1+\tilde{\sigma}^2}\frac{(\chi^2+1)^2(\chi^2-1)}{\chi^4+1}\frac{\tilde{\phi}^4}{|\xi(\tilde{\phi})|^2}\right)\,n\,\tilde{d}\,\bar{z}}\,
e^{8\,\frac{\tilde{\sigma}^2}{(1+\tilde{\sigma}^2)^2}\frac{(\chi^2-1)^2(\chi^2+1)}{\chi^4+1}\frac{\tilde{\phi}^4}{|\xi(\tilde{\phi})|^2}\,n\,\tilde{d}}\nonumber\\
&\left.\left.
e^{-2\,i\,\frac{\tilde{\sigma}^2}{1+\tilde{\sigma}^2}\,\frac{\chi^2-1}{\chi^4+1}\frac{\tilde{\phi}^2}{|\xi(\tilde{\phi})|^2}\,\tilde{d}\,n\,z_0}
e^{4\,i\,\frac{\tilde{\sigma}^2}{1+\tilde{\sigma}^2}\,\frac{\chi^3(\chi^2-1)}{\chi^4+1}\frac{\tilde{\phi}^2}{|\xi(\tilde{\phi})|^2}\,\tilde{d}\,n\,\bar{z}}
\right]\right|.
\end{align}

We now want to optimize the expression \eqref{calculation:final:comb:square:phase:appendix} with respect to $\bar{z}$. It is even more difficult to do so analytically in this case. Therefore, we resort to use the approximation $\chi\approx1+\delta_1$, and obtain
\begin{align}\label{calculation:final:approximate:comb:square:phase:appendix}
\tilde{\Delta}_{\textrm{p}}\approx&(1-\delta_1^2)\frac{
e^{-\frac{1}{8}(1+\tilde{\sigma}^2)(1-2\delta_1)\left(1+16\frac{\tilde{\phi}^4}{(1+\tilde{\sigma}^2)^2}\right)\,\frac{\bar{z}^2}{|\xi(\tilde{\phi})|^2}}\,
e^{-8\,\delta_1\,\frac{\tilde{\phi}^4}{(1+\tilde{\sigma}^2)}\frac{z_0}{|\xi(\tilde{\phi})|^2}\,\bar{z}}
e^{-8\delta_1^2\,\frac{\tilde{\phi}^4}{(1+\tilde{\sigma}^2)}\,\frac{z_0^2}{|\xi(\tilde{\phi})|^2}}
}
{\vartheta_3\bigl(0,\exp\bigl[-\frac{1}{2}\,\frac{\tilde{\sigma}^2}{1+\tilde{\sigma}^2}\tilde{d}^2\bigr]\bigr)}\nonumber\\
&\left|\sum_{n\in\mathbb{Z}}
\left[e^{-\frac{1}{2}\frac{\tilde{\sigma}^2}{1+\tilde{\sigma}^2}\left(1+\tilde{\sigma}^2\delta_1^2-32\frac{\tilde{\sigma}^2}{(1+\tilde{\sigma}^2)^2}\,\delta_1^2\,\frac{\tilde{\phi}^4}{|\xi(\tilde{\phi})|^2}-32\,i\,\frac{\tilde{\sigma}^2}{1+\tilde{\sigma}^2}\,\delta_1^2\,\frac{\tilde{\phi}^2}{|\xi(\tilde{\phi})|^2}\right)\,n^2\,\tilde{d}^2}e^{-\tilde{\sigma}^2\,\delta_1\,\left(1-64\frac{1}{1+\tilde{\sigma}^2}\,\delta_1\,\frac{\tilde{\phi}^4}{|\xi(\tilde{\phi})|^2}\right)\,n\,\tilde{d}\,\bar{z}}\,
e^{32\,\frac{\tilde{\sigma}^2}{(1+\tilde{\sigma}^2)^2}\,\delta_1^2\,\frac{\tilde{\phi}^4}{|\xi(\tilde{\phi})|^2}\,n\,\tilde{d}}\right.\right.\nonumber\\
&\left.\left.
e^{-2\,i\,\frac{\tilde{\sigma}^2}{1+\tilde{\sigma}^2}\,\delta_1\,\frac{\tilde{\phi}^2}{|\xi(\tilde{\phi})|^2}\,\tilde{d}\,n\,z_0}
e^{4\,i\,\frac{\tilde{\sigma}^2}{1+\tilde{\sigma}^2}\,\delta_1\,\frac{\tilde{\phi}^2}{|\xi(\tilde{\phi})|^2}\,\tilde{d}\,n\,\bar{z}}
\right]\right|.
\end{align}
At this point we observe that we have obtained our results for a quadratic phase $\psi(\omega)$ which has its minimum at $\omega=0$. In the dimensionless coordinates such minimum has been shifted to $z=-z_0$. Due to this (artificial) initial choice, we have the appearance of $z_0$ in the exponentials in \eqref{calculation:final:approximate:comb:square:phase:appendix}.

In the following, we choose to rescale such choice of minimum to $\omega=\omega_0$ (or, equivalently, $z=z_0$) and to study how deviations from this point affect the result \eqref{calculation:final:approximate:comb:square:phase:appendix}. This means that we can effectively replace $z_0$ with $\delta z_0$, assuming therefore that $\delta z_0=0$ represents the -- now reference -- case of the quadratic phase centered around $\omega=\omega_0$.

Therefore, we can rewrite \eqref{calculation:final:approximate:comb:square:phase:appendix} as
\begin{align}\label{calculation:final:approximate:comb:square:phase:final:reshited:appendix}
\tilde{\Delta}_{\textrm{p}}\approx&(1-\delta_1^2)\frac{
e^{-\frac{1}{8}(1+\tilde{\sigma}^2)(1-2\delta_1)\left(1+16\frac{\tilde{\phi}^4}{(1+\tilde{\sigma}^2)^2}\right)\,\frac{\bar{z}^2}{|\xi(\tilde{\phi})|^2}}\,
e^{-8\,\delta_1\,\frac{\tilde{\phi}^4}{(1+\tilde{\sigma}^2)}\frac{\delta z_0}{|\xi(\tilde{\phi})|^2}\,\bar{z}}
e^{-8\delta_1^2\,\frac{\tilde{\phi}^4}{(1+\tilde{\sigma}^2)}\,\frac{(\delta z_0)^2}{|\xi(\tilde{\phi})|^2}}
}
{\vartheta_3\bigl(0,\exp\bigl[-\frac{1}{2}\,\frac{\tilde{\sigma}^2}{1+\tilde{\sigma}^2}\tilde{d}^2\bigr]\bigr)}\nonumber\\
&\left|\sum_{n\in\mathbb{Z}}
\left[e^{-\frac{1}{2}\frac{\tilde{\sigma}^2}{1+\tilde{\sigma}^2}\left(1+\tilde{\sigma}^2\delta_1^2-32\frac{\tilde{\sigma}^2}{(1+\tilde{\sigma}^2)^2}\,\delta_1^2\,\frac{\tilde{\phi}^4}{|\xi(\tilde{\phi})|^2}-32\,i\,\frac{\tilde{\sigma}^2}{1+\tilde{\sigma}^2}\,\delta_1^2\,\frac{\tilde{\phi}^2}{|\xi(\tilde{\phi})|^2}\right)\,n^2\,\tilde{d}^2}e^{-\tilde{\sigma}^2\,\delta_1\,\left(1-64\frac{1}{1+\tilde{\sigma}^2}\,\delta_1\,\frac{\tilde{\phi}^4}{|\xi(\tilde{\phi})|^2}\right)\,n\,\tilde{d}\,\bar{z}}\,
e^{32\,\frac{\tilde{\sigma}^2}{(1+\tilde{\sigma}^2)^2}\,\delta_1^2\,\frac{\tilde{\phi}^4}{|\xi(\tilde{\phi})|^2}\,n\,\tilde{d}}\right.\right.\nonumber\\
&\left.\left.
e^{-2\,i\,\frac{\tilde{\sigma}^2}{1+\tilde{\sigma}^2}\,\delta_1\,\frac{\tilde{\phi}^2}{|\xi(\tilde{\phi})|^2}\,\tilde{d}\,n\,\delta z_0}
e^{4\,i\,\frac{\tilde{\sigma}^2}{1+\tilde{\sigma}^2}\,\delta_1\,\frac{\tilde{\phi}^2}{|\xi(\tilde{\phi})|^2}\,\tilde{d}\,n\,\bar{z}}
\right]\right|.
\end{align}
Then, we also consider only the cases where $|\delta z_0|\sim\mathcal{O}(1)$. Clearly, when $\delta z_0<0$ the parabola is centered towards the end of the wavepacket, while when $\delta z_0>0$ the parabola is centered towards the front of the wavepacket. With this in mind, we note that when $\delta_1=0$, optimization of \eqref{calculation:final:approximate:comb:square:phase:final:reshited:appendix} would trivially give $\bar{z}_{\textrm{opt}}\approx0$. Therefore, we expect that $\bar{z}_{\textrm{opt}}\approx\mu\delta_1$ for an appropriate coefficient $\mu$ that depends on all of the other parameters as such that $\mu\delta_1\ll1$.
\begin{align}\label{calculation:final:approximate:comb:square:phase:final:reshited:new:appendix}
\tilde{\Delta}_{\textrm{p}}\approx&(1-\delta_1^2)\frac{
e^{-\frac{\mu^2}{8}\,\delta_1^2\,(1+\tilde{\sigma}^2)(1-2\delta_1)\left(1+16\frac{\tilde{\phi}^4}{(1+\tilde{\sigma}^2)^2}\right)\,\frac{1}{|\xi(\tilde{\phi})|^2}}\,
e^{-8\,\mu\,\delta_1^2\,\frac{\tilde{\phi}^4}{(1+\tilde{\sigma}^2)}\frac{\delta z_0}{|\xi(\tilde{\phi})|^2}}
e^{-8\delta_1^2\,\frac{\tilde{\phi}^4}{(1+\tilde{\sigma}^2)}\,\frac{(\delta z_0)^2}{|\xi(\tilde{\phi})|^2}}
}
{\vartheta_3\bigl(0,\exp\bigl[-\frac{1}{2}\,\frac{\tilde{\sigma}^2}{1+\tilde{\sigma}^2}\tilde{d}^2\bigr]\bigr)}\nonumber\\
&\left|\sum_{n\in\mathbb{Z}}
\left[e^{-\frac{1}{2}\frac{\tilde{\sigma}^2}{1+\tilde{\sigma}^2}\left(1+\tilde{\sigma}^2\delta_1^2-32\frac{\tilde{\sigma}^2}{(1+\tilde{\sigma}^2)^2}\,\delta_1^2\,\frac{\tilde{\phi}^4}{|\xi(\tilde{\phi})|^2}-32\,i\,\frac{\tilde{\sigma}^2}{1+\tilde{\sigma}^2}\,\delta_1^2\,\frac{\tilde{\phi}^2}{|\xi(\tilde{\phi})|^2}\right)\,n^2\,\tilde{d}^2}e^{-\mu\,\tilde{\sigma}^2\,\delta_1^2\,\left(1-64\frac{1}{1+\tilde{\sigma}^2}\,\delta_1\,\frac{\tilde{\phi}^4}{|\xi(\tilde{\phi})|^2}\right)\,n\,\tilde{d}}\,
e^{32\,\frac{\tilde{\sigma}^2}{(1+\tilde{\sigma}^2)^2}\,\delta_1^2\,\frac{\tilde{\phi}^4}{|\xi(\tilde{\phi})|^2}\,n\,\tilde{d}}\right.\right.\nonumber\\
&\left.\left.
e^{-2\,i\,\frac{\tilde{\sigma}^2}{1+\tilde{\sigma}^2}\,\delta_1\,\frac{\tilde{\phi}^2}{|\xi(\tilde{\phi})|^2}\,\tilde{d}\,n\,\delta z_0}
e^{4\,\mu\,i\,\frac{\tilde{\sigma}^2}{1+\tilde{\sigma}^2}\,\delta_1\,\frac{\tilde{\phi}^2}{|\xi(\tilde{\phi})|^2}\,\tilde{d}\,n}
\right]\right|.
\end{align}
We see that all of the coefficients in the exponentials that contain the linear $n\,\tilde{d}$ are small. If we write the expansion of each exponential explicitly, all terms that are proportional to $(n\,\tilde{d})^{2k+1}$ vanish identically due to the sum running over all integers. Therefore, the expansion of such exponentials will leave only terms that are proportional to $(n\,\tilde{d})^{2k}$. Assuming that it is appropriate to retain only such terms, we can therefore expand to second order all exponentials that are linear $n\,\tilde{d}$, and eliminate coefficients that are of higher order in each term of the perturbative expansions, which leaves us with the expression
\begin{align}\label{calculation:final:approximate:comb:square:phase:final:reshited:new:to:be:optimizedappendix}
\tilde{\Delta}_{\textrm{p}}\approx&(1-\delta_1^2)
e^{-\frac{\mu^2}{8}\,\tilde{\sigma}^2\,\delta_1^2}\,e^{-8\,\mu\,\delta_1^2\,\frac{\tilde{\phi}^4}{\tilde{\sigma}^2}\,\delta z_0}\,e^{-8\delta_1^2\,\frac{\tilde{\phi}^4}{\tilde{\sigma}^2}\,(\delta z_0)^2}
\left|\frac{\vartheta_3\left(0,e^{-\frac{1}{2}\,\left(1+\tilde{\sigma}^2\delta_1^2-32\frac{\delta_1^2\,\tilde{\phi}^4}{\tilde{\sigma}^2}-32\,i\,\delta_1^2\,\tilde{\phi}^2\right)\,\tilde{d}^2}\right)}
{\vartheta_3\bigl(0,\exp\bigl[-\frac{1}{2}\,\tilde{d}^2\bigr]\bigr)}\right.\nonumber\\
&\left.\times\left[1+2\,\Theta\,\tilde{d}^2\,\delta_1^2\,\sum_{n=1}^{+\infty}\cosh^{-2}\left(\frac{1}{2}\,(n-1/2)\left(1+\tilde{\sigma}^2\delta_1^2-32\frac{\delta_1^2\,\tilde{\phi}^4}{\tilde{\sigma}^2}-32\,i\,\delta_1^2\,\tilde{\phi}^2\right)\,\tilde{d}^2\right)\right]\right|.
\end{align}
Here we have introduced the quantity $\Theta$ defined as
\begin{align}
\Theta\approx&\mu^2\tilde{\sigma}^4\,\delta_1^2+\left[1024\frac{1}{\tilde{\sigma}^2}\tilde{\phi}^6\delta_1^2-4(\delta z_0)^2-16\mu^2-32\mu\tilde{\phi}^2\delta_1^2+2\,i\,\tilde{\phi}^2\,\delta z_0-4\,\mu^2\,i\,\tilde{\phi}^2\delta_1-64\,i\,\frac{1}{\tilde{\sigma}^2}\tilde{\phi}^4\delta_1\delta z_0\right.\nonumber\\
&\left.+128\,\mu\,i\frac{1}{\tilde{\sigma}^2}\tilde{\phi}^4\delta_1+8\mu\tilde{\phi}^2\delta z_0\right]\,\tilde{\phi}^2
\end{align}
and we have used the following properties of the Jacobi theta function:
\begin{align}
\frac{d^2}{dz^2}\vartheta_3\left(z,q\right)|_{z=0}=&-8\sum_{n=1}^{+\infty}\frac{q^{2n-1}}{(1-q^{2n-1})^2}\,\vartheta_3\left(0,q\right)\nonumber\\
\frac{d^2}{dz^2}\vartheta_3\left(z,q\right)|_{z=0}=&-q\frac{d}{dq}\vartheta_3\left(0,q\right).
\end{align}
Using again the expansion of $\vartheta_3(0,\exp\bigl[-x\bigr])$ for $x\ll1$, we can finally obtain
\begin{align}\label{calculation:final:approximate:comb:square:phase:final:that:can:be:used:appendix}
\tilde{\Delta}_{\textrm{p}}\approx&(1-\delta_1^2)
e^{-\frac{\mu^2}{8}\,\tilde{\sigma}^2\,\delta_1^2}\,e^{-8\,\mu\,\delta_1^2\,\frac{\tilde{\phi}^4}{\tilde{\sigma}^2}\,\delta z_0}\,e^{-8\delta_1^2\,\frac{\tilde{\phi}^4}{\tilde{\sigma}^2}\,(\delta z_0)^2}
\left(1-\tilde{\sigma}^2\delta_1^2+32\frac{\delta_1^2\,\tilde{\phi}^4}{\tilde{\sigma}^2}\right)\nonumber\\
&\times\left|\left[1+2\,\Theta\,\tilde{d}^2\,\delta_1^2\,\sum_{n=1}^{+\infty}\cosh^{-2}\left(\frac{1}{2}\,(n-1/2)\left(1+\tilde{\sigma}^2\delta_1^2-32\frac{\delta_1^2\,\tilde{\phi}^4}{\tilde{\sigma}^2}-32\,i\,\delta_1^2\,\tilde{\phi}^2\right)\,\tilde{d}^2\right)\right]\right|\nonumber\\
\approx&(1-\delta_1^2)
e^{-\frac{\mu^2}{8}\,\tilde{\sigma}^2\,\delta_1^2}\,e^{-8\,\mu\,\delta_1^2\,\frac{\tilde{\phi}^4}{\tilde{\sigma}^2}\,\delta z_0}\,e^{-8\delta_1^2\,\frac{\tilde{\phi}^4}{\tilde{\sigma}^2}\,(\delta z_0)^2}
\left(1-\tilde{\sigma}^2\delta_1^2+32\frac{\delta_1^2\,\tilde{\phi}^4}{\tilde{\sigma}^2}\right)\nonumber\\
&\times\left[1+2\,\tilde{d}^2\,\delta_1^2\,\Re\left(\Theta\,\sum_{n=1}^{+\infty}\cosh^{-2}\left(\frac{1}{2}\,(n-1/2)\left(1+\tilde{\sigma}^2\delta_1^2-32\frac{\delta_1^2\,\tilde{\phi}^4}{\tilde{\sigma}^2}-32\,i\,\delta_1^2\,\tilde{\phi}^2\right)\,\tilde{d}^2\right)\right)\right].
\end{align}
It is still difficult to optimize the expression \eqref{calculation:final:approximate:comb:square:phase:final:that:can:be:used:appendix} with respect to $\mu$. At this point, we can also use the rough estimate $\sum_{n=1}^{+\infty}\cosh^{-2}(n\,x)\approx \zeta/x$ for $x\ll1$, with $\zeta$ a numerical constant that can be determined numerically if necessary, or explicitly if known properties of sums of inverse hyperbolic functions can be exploited.  This rough estimate can be justified by saying that as long as $nx\leq1$, then $\cosh^{-2}(n\,x)\approx1$. As soon as $nx>1$, we have that $\cosh^{-2}(n\,x)$ decreases exponentially. This, we say that $\cosh(n\,x)\sim1$ for $n\leq x$, and $\zeta$ corrects for the error in this assumption and the tail of contributions that come from the exponential decay. We expect that $\zeta$ will not have values that impact significantly the claims below and, in any case, that $\zeta\geq1$ since we are overestimating the contributions to the sum for $n<x$ and underestimating them in the exponentially decaying tail of $n>x$. 

Therefore, since $\tilde{d}^2\,\delta_1^2\,\Theta\ll1$, we have 
\begin{align}\label{calculation:final:approximate:comb:square:phase:final:that:can:be:used:good:appendix}
\tilde{\Delta}_{\textrm{p}}\approx&(1-\delta_1^2)
e^{-\frac{\mu^2}{8}\,\tilde{\sigma}^2\,\delta_1^2}\,e^{-8\,\mu\,\delta_1^2\,\frac{\tilde{\phi}^4}{\tilde{\sigma}^2}\,\delta z_0}\,e^{-8\delta_1^2\,\frac{\tilde{\phi}^4}{\tilde{\sigma}^2}\,(\delta z_0)^2}
\left(1-\frac{1}{2}\tilde{\sigma}^2\delta_1^2+16\frac{\delta_1^2\,\tilde{\phi}^4}{\tilde{\sigma}^2}\right)\,\left|\left[1+2\,\zeta\,\tilde{d}^2\,\Theta\,\delta_1^2\right]\right|\nonumber\\
\approx&(1-\delta_1^2)
e^{-\frac{\mu^2}{8}\,\tilde{\sigma}^2\,\delta_1^2}\,e^{-8\,\mu\,\delta_1^2\,\frac{\tilde{\phi}^4}{\tilde{\sigma}^2}\,\delta z_0}\,e^{-8\delta_1^2\,\frac{\tilde{\phi}^4}{\tilde{\sigma}^2}\,(\delta z_0)^2}
\left(1-\frac{1}{2}\tilde{\sigma}^2\delta_1^2+16\frac{\delta_1^2\,\tilde{\phi}^4}{\tilde{\sigma}^2}\right)\,\left[1+2\,\zeta\,\tilde{d}^2\Re\Theta\,\delta_1^2\right].
\end{align}
Optimization of \eqref{calculation:final:approximate:comb:square:phase:final:that:can:be:used:good:appendix} over $\mu$ gives us the optimal value $\mu_{\textrm{opt}}$ with expression
\begin{align}\label{optimal:mu:preliminary:appendix}
\mu_{\textrm{opt}}=32\frac{(4\,\zeta\,\tilde{d}^2\,\tilde{\sigma}^2-1)\delta z_0-16\,\zeta\,\tilde{d}^2\,\tilde{\sigma}^2\delta_1^2}{\tilde{\sigma}^2-32\,\zeta\,\tilde{d}^2\,\tilde{\sigma}^4\delta_1^2+16\,\zeta\,\tilde{d}^2\,\tilde{\phi}^2}\frac{\tilde{\phi}^4}{\tilde{\sigma}^2}.
\end{align}
This can be further simplified by noting that we expect $\zeta\,\tilde{d}^2\tilde{\sigma}^2\gg1\gg\delta_1^2$, and by noting that $\tilde{\sigma}^2\gg1\gg\tilde{d}^2\,\tilde{\sigma}^4\delta_1^2$, since $\tilde{d}^2\,\tilde{\sigma}^4\delta_1^2\ll1$ in order for the perturbation theory to apply. Therefore, we have
\begin{align}\label{optimal:mu:final:appendix}
\mu_{\textrm{opt}}=128\,\zeta\,\tilde{d}^2\,\tilde{\phi}^4\,\frac{\delta z_0-4\,\delta_1^2}{\tilde{\sigma}^2+16\,\zeta\,\tilde{d}^2\,\tilde{\phi}^2}.
\end{align}
Note that $128\,\zeta\,\tilde{d}^2\,\tilde{\phi}^4\delta_1^2\ll1$, which can be seen during the derivation of \eqref{optimal:mu:preliminary:appendix}. 

We can study the two opposite cases of interest: (i) where $\tilde{\phi}\sim\mathcal{O}(1)$, and (ii) where $\tilde{\phi}\gg1$:
\begin{itemize}
\item[\textit{(i)}] In the first case, when $\tilde{\phi}\sim\mathcal{O}(1)$, we always have that
\begin{align}\label{calculation:final:approximate:comb:square:phase:final:case:one:appendix}
\tilde{\Delta}_{\textrm{p,opt}}^{\textrm{Co}}\approx&1-\delta_1^2-\frac{1}{2}\tilde{\sigma}^2\delta_1^2.
\end{align}
Note that the result here is independent on $\delta z_0$.
\item[\textit{(ii)}] In the second case, we have two possibilities. When $\delta z_0\sim\nu\,\delta_1^2$, with $\nu$ a constant that is not large then we have $\mu_{\textrm{opt}}\approx\nu\,\delta_1^2$ and $z_{\textrm{opt}}\approx0$.  Therefore, we have $\Theta\approx1024\frac{\tilde{\phi}^4}{\tilde{\sigma}}\,\delta_1^2$  we can use \eqref{calculation:final:approximate:comb:square:phase:final:that:can:be:used:good:appendix} to find the optimal overlap 
\begin{align}\label{calculation:final:approximate:comb:square:phase:final:case:two:appendix}
\tilde{\Delta}_{\textrm{p,opt}}^{\textrm{Co}}\approx&1-\delta_1^2-\frac{1}{2}\tilde{\sigma}^2\delta_1^2+16\frac{\tilde{\phi}^4}{\tilde{\sigma}^2}\delta_1^2.
\end{align}
\end{itemize}
We can also have $\delta z_0\geq1$, and therefore
\begin{align}\label{optimal:mu:final:second:appendix}
\mu_{\textrm{opt}}=8\tilde{\phi}^2\,\frac{\delta z_0}{1+\Sigma},
\end{align}
where $\Sigma:=\tilde{\sigma}^2/(16\,\zeta\,\tilde{d}^2\,\tilde{\phi}^2)$.
Finally, we find
\begin{align}\label{calculation:final:approximate:comb:square:phase:final:case:three:appendix}
\tilde{\Delta}_{\textrm{p,opt}}^{\textrm{Co}}\approx&1-\delta_1^2-\frac{1}{2}\tilde{\sigma}^2\delta_1^2+16\frac{\tilde{\phi}^4}{\tilde{\sigma}^2}\delta_1^2
-8\frac{\tilde{\phi}^2}{\tilde{d}^2\,(1+\Sigma)^2}\left[\tilde{\sigma}^4\tilde{\phi}^2+8(1+\Sigma)+\tilde{\phi}^2(1+\Sigma)^2+\zeta\,\tilde{d}^2\,\tilde{\sigma}^2\,(1+\Sigma)^2\right.\nonumber\\
&\left.+256\,\tilde{d}^2\,\tilde{\sigma}^2\,\tilde{\phi}^2-16\zeta\,\tilde{d}^2\,\tilde{\sigma}^2\,\tilde{\phi}^4\,(1+\Sigma)\right]\,\delta z_0^2\,\delta_1^2.
\end{align}
It is clear that, when $\delta z_0\sim\nu\,\delta_1^2$, then we can effectively set the terms proportional to $\delta z_0$ in \eqref{calculation:final:approximate:comb:square:phase:final:case:three:appendix} to zero and, as expected, the result \eqref{calculation:final:approximate:comb:square:phase:final:case:three:appendix} reduces to \eqref{calculation:final:approximate:comb:square:phase:final:case:two:appendix}.

%---------------------------------------------------------------------------------------------------%
\end{document}